%% file: arxiv_version.tex
\documentclass[runningheads]{llncs}
\usepackage{graphicx}
\usepackage{caption}
\usepackage{subcaption}
\usepackage{todonotes,soul}

\input{macros}
\begin{document}
\title{MANTRA: Temporal Betweenness Centrality Approximation through Sampling}
\titlerunning{Temporal Betweenness Centrality Approximation through Sampling}
%
%

%
\authorrunning{}
%
\author{Antonio Cruciani}
\institute{Gran Sasso Science Institute,
 L’Aquila, Italy\\ \email{antonio.cruciani@gssi.it}}

\maketitle              
\input{trunk/abstract}

\input{trunk/introduction}
\input{trunk/related_works}

\input{trunk/background_and_notation}
\input{trunk/approximation_algorithms}
\input{trunk/experiments}
\input{trunk/conclusions}
\bibliographystyle{splncs04}
\bibliography{references}
\newpage
\appendix
\begin{center}
    \LARGE{\textbf{Appendix}}
\end{center}
\input{trunk/appendix}
\end{document}

%% file: macros.tex
\newcommand{\Prob}[1]{{\textbf{Pr}} \left( #1 \right)}
\newcommand{\Var}[1]{{\textbf{Var}} \left[ #1 \right]}

\newcommand{\Expec}[1]{\textbf{E} \left[ #1 \right]}
\newcommand{\CExpec}[2]{\textbf{E}_{#2}\left[ #1\right]}

\newcommand{\afuncc}[2]{f^{(\star)}_{#2}({#1}) }
\newcommand{\afunc}[2]{\widetilde{b}^{(\star)}_{#2}(#1) }

\newcommand{\gbet}[1]{\texttt{b}^{(\star)}_{#1} }

\newcommand{\agbet}[1]{\widetilde{\texttt{b}}^{(\star)}_{#1} }

\newcommand{\sopt}[1]{ \Gamma_{#1}^{(\star)} }
\newcommand{\dopt}{\mathbb{TP}^{(\star)}}

\newcommand{\optpath}{$(\star)$-temporal path }
\newcommand{\optpaths}{$(\star)$-temporal paths }

\newcommand{\calS}{\mathcal{S}}

\newcommand{\calU}{\mathcal{U}}
\newcommand{\bigO}{\mathcal{O}}

\newcommand{\calG}{\mathcal{G}}
\newcommand{\calB}{\mathcal{B}}
\newcommand{\calE}{\mathcal{E}}
\newcommand{\calF}{\mathcal{F}}

\newcommand{\calD}{\mathcal{D}}

\usepackage[linesnumbered,ruled,vlined,boxed]{algorithm2e}
\newlength{\commentWidth}
\let\oldnl\nl
\newcommand{\nonl}{\renewcommand{\nl}{\let\nl\oldnl}}
\DontPrintSemicolon
\SetKwInOut{Input}{Input}\SetKwInOut{Output}{Output}

\newcommand{\argmax}{\texttt{argmax}}
\newcommand{\ssbe}{\texttt{\textbf{rtb}}}

\newcommand{\onbra}{\texttt{\textbf{ob}}}
\newcommand{\srtp}{\texttt{\textbf{trk}}}

\usepackage{xspace}
\usepackage{booktabs}

\usepackage{hyperref}
\usepackage{mathtools}
\usepackage{amsmath}
\usepackage{amsfonts}
\usepackage{bm}
\usepackage{dsfont}
\usepackage{upgreek}

%% file: trunk/abstract.tex
\begin{abstract}
We present MANTRA, a framework for approximating the temporal betweenness centrality of all nodes in a temporal graph. Our method can compute probabilistically guaranteed high-quality temporal betweenness estimates (of nodes and temporal edges) under all the feasible temporal path optimalities, presented in the work of Bu{\ss} et al. (KDD, 2020). We provide a sample-complexity analysis of our method and speed up the temporal betweenness computation using a state-of-the-art progressive sampling approach based on Monte Carlo Empirical Rademacher Averages. Additionally, we provide an efficient sampling algorithm to approximate the temporal diameter, average path length, and other fundamental temporal graph characteristic quantities within a small error $\varepsilon$ with high probability. The running time of such approximation algorithm is $\tilde{\bigO}(\frac{\log n}{\varepsilon^2}\cdot |\calE|)$, where $n$ is the number of nodes and $|\calE|$ is the number of temporal edges in the temporal graph. We support our theoretical results with an extensive experimental analysis on several real-world networks and provide empirical evidence that the MANTRA framework improves the current state of the art in speed, sample size, and required space while maintaining high accuracy of the temporal betweenness estimates. 
\end{abstract}

%% file: trunk/introduction.tex
\section{Introduction}
Centrality measures are fundamental notions for evaluating the importance of nodes in networks, used in network analysis and graph theory. A centrality measure assigns real values to all the nodes, in such a way that the values are monotonously dependent of the nodes' importance, i.e., more important nodes should have higher centrality scores. Computing the \emph{betweenness centrality} is arguably one of the most important tasks in graph mining and network analysis. It finds application in several fields including social network analysis~\cite{Tang_2010}, routing~\cite{Daly_2007}, machine learning~\cite{Simsek_2008}, and neuroscience~\cite{Van_2010}. The betweenness of a node in a graph indicates how often this node is visited by a shortest path. High betweenness nodes are usually considered to be important in the network. Brandes' algorithm~\cite{Brandes_2001}, is the best algorithm to compute the exact centrality scores of every node in $\bigO(n\cdot m)$ time and $\bigO(n+ m)$ space where $n$ and $m$ are the number of nodes, and edges of a given graph $G=(V,E)$, respectively. Unfortunately, this algorithm quickly becomes impractical on nowadays' networks with billions of nodes and edges. Moreover, there is a theoretical evidence, in form of several conditional lower bounds results~\cite{Abboud_2015}, for believing that a faster algorithm cannot exists, even for \emph{approximately} computing the betweenness. A further challenge, is that modern real-world networks are also dynamic or temporal, i.e., they change over time. Temporal networks can be informally described as edge-labeled graphs in which each label indicates the time instant in which the underlying edge appears in the network. A great variety of both modern and traditional networks can be naturally modeled as temporal graphs. 
Furthermore, there are numerous real-world applications for which studying temporal networks offers unique perspectives. This is especially evident when examining data that evolves over time, for example social networks interactions, information/infection spreading, subgraph patterns, detecting communities, and clustering networks. In the context of these challenges it is, thus, essential to consider temporal variants of the most important centrality notions, alongside algorithms for computing them, that have an excellent scaling behavior. In this work, we focus on the temporal version of the \emph{betweenness centrality}, that similarly to static networks, it seeks to pinpoint nodes that are traversed by a significant number of optimal (temporal) paths. 
Bu{\ss} et al.~\cite{Buss_2020,Rymar_2021} gave several definitions of the \emph{temporal betweenness} as a temporal counterpart of the \emph{betweenness centrality}, characterized their computational complexity, and provided polynomial time algorithms to compute these temporal centrality measures. However, these algorithms turn out to be impractical, even for medium size networks. Thus, it is reasonable to consider approximation algorithms that can efficiently compute the centrality values of the nodes up to some small error. In this work, we follow the approach of Santoro et al.~\cite{Santoro_2022}, and we provide a set of approximation algorithms for all the temporal betweenness variants in~\cite{Buss_2020}. 
\paragraph{Contributions.} We propose {\sc MANTRA} (\textit{teMporAl betweeNnes cenTrality thRough sAmpling}), a rigorous framework for the approximation of the temporal betweenness of all the vertices and temporal edges in large temporal graphs. In particular, we present the following results: (1) We extend the state-of-the-art estimator~\cite{Santoro_2022} to all the feasible temporal betweenness centrality variants for nodes and temporal edges (Section~\ref{sec:estimators}). In addition, we propose two alternative unbiased estimators for such centrality measure on temporal graphs\footnote{Due to space constraints, we refer to the additional materials for the temporal betweenness on temporal edges and for the definition of the other estimators.}; (2) We derive new bounds on the sufficient number of samples to approximate the temporal betweenness centrality for all nodes\footnote{The sample complexity analysis holds also for the temporal edges.} (Section~\ref{sec::ss_bound}), that are governed by three key quantities of the temporal graph, such as the \emph{temporal vertex diameter}, \emph{average temporal path length}, and the \emph{maximum variance} of the temporal betweenness centrality estimators. Moreover, this result solves an open problem in~\cite{Pellegrina_2021,Pellegrina_2023} on whether the sample complexity bounds for the static betweenness can be \emph{efficiently} extended to temporal graphs. As a consequence, it significantly improves on the state-of-the-art results for the temporal betweenness estimation process~\cite{Santoro_2022}. Additionally, our analysis of sample complexity presents further challenges regarding the efficient computation of the three quantities upon which the bounds for the necessary sample size depend;
(3) We propose a novel algorithm to efficiently estimate the key quantities of interests in (2) that uses a mixed approach based on \emph{sampling} and \emph{counting} (Section~\ref{sec::diam_apx}). The time complexity of our approach is $\tilde{\bigO}(\frac{\log n}{\varepsilon^2}|\calE|)$, while the space complexity is $\bigO(n+|\calE|)$. We provide an estimate on the sample size needed to achieve a good estimates up to a small error bound. More precisely, we prove that $r = \Theta(\frac{\log n}{\varepsilon^2})$ sample nodes are sufficient to estimate, with probability at least $1-1/n^2$: (i) the temporal diameter $D^{(\star)}$ with error bounded by $\frac{\varepsilon}{\zeta^{(\star)}}$; (ii) the average temporal path length $\rho^{(\star)}$ with error bounded by $\varepsilon\frac{D^{(\star)}}{\zeta^{(\star)}}$; and, (iii) the temporal connectivity rate $\zeta^{(\star)}$ (see Section~\ref{sec::diam_apx} for the formal definition) with error bounded by $\varepsilon$; (4) We define {\sc MANTRA}, a progressive sampling algorithm that uses an advanced tool from statistical learning theory, namely \emph{Monte Carlo Empirical Rademacher Averages}~\cite{Barlett_2003} and the above results (e.g. (1-3)) to provide a high quality approximation of the temporal betweenness (Section~\ref{sec::mantra_framework}). MANTRA's output is a function of two parameters: $\varepsilon \in (0,1)$ controlling the approximation's accuracy, and $\delta\in (0,1)$ controlling the confidence of the computed approximation. Our novel approach improves on ONBRA~\cite{Santoro_2022} (i.e., the state-of-the-art algorithm) in terms of running time, sample size, and allocated space; and, (5) We support our theoretical analysis with an extensive experimental evaluation (Section~\ref{sec::experiments}), in which we compare MANTRA with ONBRA.

%% file: trunk/related_works.tex
\section{Related Work}
The literature on betweenness centrality being vast, we restrict our attention to approaches that are closest to ours. Thus, we focus on betweenness centrality on temporal graphs. 
Tsalouchidou et al.~\cite{Tsalouchidou_2020}, extended the well-known Brandes algorithm~\cite{Brandes_2001} to allow for distributed computation of betweenness in temporal graphs. Specifically, they studied shortest-fastest paths, considering the bi-objective of shortest length and shortest duration. Bu{\ss} et al.~\cite{Buss_2020,Rymar_2021} analysed the temporal betweenness centrality considering several temporal path optimality criteria, such as shortest (foremost), foremost, fastest, and prefix-foremost, along with their computational complexities. They showed that, when considering paths with increasing time labels, the foremost and fastest temporal betweenness variants are $\#P$-hard, while the shortest and shortest foremost ones can be computed in $O(n^3 \cdot |T|^2)$, and the prefix-foremost one in $O(n\cdot |\calE|\cdot \log |\calE|)$. Here $\calE$ is the set of temporal edges, and $T$ is the set of unique time stamps. Santoro et al.~\cite{Santoro_2022} provided ONBRA, the first sampling-based approximation algorithm for one variant of the temporal betweenness centrality. The input to ONBRA is a temporal graph, a confidence value $\delta\in (0,1)$, and the sample size $r$. The algorithm performs a set of $r$ truncated temporal breadth first searches between couples of nodes sampled uniformly at random and estimates the shortest temporal betweenness using the temporal equivalent of the ABRA estimator~\cite{Riondato_2018} for static networks. ONBRA's output is a function of the confidence $\delta\in (0,1)$ and the upper bound on the approximation accuracy $\xi\in (0,1)$ computed using the \emph{Empirical Bernstein Bound}~\cite{Maurer_2009}. More precisely, with probability $1-\delta$, the approximation computed by ONBRA is guaranteed to have absolute error of at most $\xi$ for each node in the temporal graph. Finally, Becker et al.~\cite{Becker_2023}, provided an efficient heuristic to approximate the temporal betweenness rankings by considering the temporal interactions among the $1$-hop neighborhood of the nodes.

%% file: trunk/background_and_notation.tex
\section{Preliminaries}\label{sec:bg}

\subsubsection{Temporal Graphs, and Paths.} 
A directed \emph{temporal graph} is an ordered tuple $\calG= (V,\calE)$ where $\calE = \{(u,v,t):u,v,\in V \land t\in T\subseteq \mathbb{N}\}$ is the set of \emph{temporal edges}\footnote{The value $T$ denotes the \emph{life-time} of the temporal graph, and, without loss of generality for our purposes, we assume that, for any $t\in T$, there exists at least one temporal arc at that time and without loss of generality we assume $T= [1,|T|]$.}.  
Given two nodes $s$ and $z$, a \emph{temporal path} $tp_{sz}\subseteq V\times T$ is a (unique) sequence of time-respecting vertex appearances $tp_{sz}=\left((u_1,t_1),(u_2,t_2),\dots , (u_k,t_k)\right)$ such that $u_1 = s$, $u_k = z$, and $t_i<t_{i+1}$ for all $1\leq i\leq k-1$. 
The vertex appearances $(u_1,t_1)$ and $(u_k,t_k)$ are called \emph{endpoints} of $tp_{sz}$ and the temporal nodes in $\textbf{Int}(tp_{sz}) = tp_{sz}\setminus \{(u_1,t_1),(u_k,t_k)\}$ are called  \emph{internal vertex appearances} of $tp_{sz}$. Unlike paths on static graphs, in the temporal setting there are several concepts of optimal paths (e.g., \emph{shortest}, \emph{foremost}, \emph{fastest})~\cite{Xuan_2003,Buss_2020,Rymar_2021}. Moreover, as for the static betweenness, the task of computing the desired centrality measure boils down to the ability of efficiently \emph{counting} the overall number of optimal paths. Unfortunately, it has been already shown that such task turns out to be \#P-Hard for some temporal path optimalities (e.g. foremost, fastest)~\cite{Buss_2020,Rymar_2021}. Hope is left for the shortest (and all its variants) an the prefix foremost temporal paths. 
We formally describe those that can be efficiently counted. 
\begin{definition}\label{def::tps}
	Given a temporal graph $\calG$, and two nodes $s,z\in V$. Let $tp_{sz}$ be a temporal path from $s$ to $z$, then $tp_{sz}$ is said to be:
	(1) \emph{Shortest} (sh) if there is no $tp_{sz}'$ such that $|tp_{sz}'|<|tp_{sz}|$; (2) \emph{Shortest-Foremost} (sfm) if there is no $tp_{sz}'$ that has an earlier arrival time in $z$ than $tp_{sz}$ and has minimum length in terms of number of hops from $s$ to $z$; and, (3) \emph{Prefix-Foremost} (pfm) if $tp_{sz}$ is foremost and every prefix $tp_{sv}$ of $tp_{sz}$ is foremost as well.
\end{definition}
To denote the different type of temporal paths
we use the same notation of Bu{\ss} et al. \cite{Buss_2020}. More precisely, we use the term ``$(\star)$-optimal'' temporal path, where $(\star)$ denotes the type. 
Furthermore, we denote the set of \emph{all} \optpaths between two nodes $s$ and $z$ as $\sopt{sz}$ and we let $\dopt_{\calG}$ to be the union of all the $\sopt{sz}$'s, for all pairs $(s,z)\in V\times V$ of distinct nodes. In this work, we will heavily rely on two temporal graphs characteristic quantities, namely the \emph{temporal (vertex) diameter} and the \emph{average temporal path length}. Formally, given a temporal graph $\calG= (V,\calE)$ we define the \emph{$(\star)$-temporal diameter} $D^{(\star)}$ and the \emph{$(\star)$-temporal vertex diameter} $VD^{(\star)}$ as the number of temporal edges and nodes in the longest $(\star)$-optimal path in $\calG$, i.e., $$D^{(\star)} = \max\left\{|tp^{(\star)}| : tp^{(\star)} \in \dopt_\calG \right\}\text{ and } VD^{(\star)}=D^{(\star)}+1$$ respectively. Finally, we refer to the average \emph{$(\star)$-temporal path length} $\rho^{(\star)}$ as the average number of internal nodes in a \emph{$(\star)$-temporal path}, i.e., $$\rho^{(\star)} = \frac{1}{n(n-1)} \sum_{s,z\in V} |\textbf{Int}(tp_{sz})| $$

\subsubsection{Temporal Betweenness Centrality.}



As previously shown, on temporal graphs, there are several notions of optimal paths. Hence, we have different notions of \emph{temporal betweenness centrality}~\cite{Buss_2020} as well. Formally, let $\calG = (V,\calE)$ be a temporal graph. For any pair $(s,z)$ of distinct nodes $(s\neq z)$, let $\sigma_{sz}^{(\star)}$ be the number of \optpaths between $s$ and $z$, and let $\sigma_{sz}^{(\star)}(v)$ be the number of the \optpaths between $s$ and $z$ that \emph{pass through} node $v$, with $s\neq v\neq z$. The normalized \emph{temporal betweenness centrality} $\gbet{v}$ of a node $v\in V$ is defined as
$$\gbet{v} =\frac{1}{n(n-1)}\sum_{\substack{s\neq v\neq z }}{\frac{\sigma_{sz}^{(\star)}(v)}{\sigma^{(\star)}_{sz}}}$$
We refer to the additional materials for the definition of the $(\star)$-temporal betweenness of the temporal edges. Whenever we use the term \optpaths we consider $(\star)$ to be one of the optimality criteria in Definition \ref{def::tps}. We observe that the average \emph{$(\star)$-temporal path length} is equal to the sum of the $(\star)$-temporal betweenness centrality over all nodes $v\in V$.
\begin{lemma}\label{lemma::diam_avg_dist}
	$\rho^{(\star)} = \sum_{v\in V}\gbet{v}$
\end{lemma} 

\subsubsection{Supremum Deviation and Empirical Rademacher Averages.}
Here we define the \emph{Supremum Deviation} (SD) and the \emph{c-samples Monte Carlo Empirical Rademacher Average} (c-MCERA). For more details about the topic we refer to the book~\cite{Shalev_2014} and to~\cite{Barlett_2003}. Let $\calD$ be a finite domain and consider a probability distribution $\pi$ over the elements of $\calD$. Let $\calF$ be a family of functions from $\calD$ to $[0,1]$, and $\calS= \{s_1,\dots,s_r\}$ be a collection of $r$ independent and identically distributed samples from $\calD$ sampled according to $\pi$. 
The SD is defined as:
$$SD(\calF,\calS) = \sup_{f\in \calF}\left|\frac{1}{r}\sum_{i\in [r]}f(s_i)- \CExpec{f}{\pi}\right|$$
The SD is the key concept of the study of empirical processes~\cite{Pollard_2012}. One way to derive probabilistic upper bounds to the SD is to use the \emph{Empirical Rademacher Averages} (ERA)~\cite{Koltchinskii_2001}.
In this work we use the state-of-the-art approach to obtain sharp probabilistic bounds on the ERA that uses Monte-Carlo estimation~\cite{Barlett_2003}. Consider a sample $\calS= \{s_1,\dots s_r\}$, for $c\geq 1$ let $\bm{\lambda}\in \{-1,1\}^{c\times r}$ be a $c\times r$ matrix of i.i.d. Rademacher random variables. The c-MCERA of $\calF$ on $\calS$ using $\bm{\lambda}$ is:
	$$R_r^c(\calF,\calS,\bm{\lambda}) = \frac{1}{c}\sum_{j=1}^{c}\sup_{f\in \calF} \frac{1}{r}\sum_{i = 1}^r \lambda_{j,i}f(s_i)$$
The c-MCERA allows to obtain sharp data-dependent probabilistic upper bounds on the SD, as they directly estimate the expected SD of sets of functions by taking into account their correlation. Moreover, they are often significantly more accurate than other methods~\cite{Pellegrina_2021,Pellegrina_2022,Pellegrina_2023}, such as the ones based on loose deterministic upper bounds to ERA~\cite{Riondato_2018}, distribution-free notions of complexity such as the Hoeffding's bound or the VC-Dimension, or other results on the variance~\cite{Maurer_2009,Santoro_2022}. 
Moreover, a key quantity governing the accuracy of the c-MCERA is the \emph{empirical wimpy variance}~\cite{Boucheron_2013} $\mathcal{W}_\calF(\calS)$, that for a sample of size $r$ is defined as $\mathcal{W}_\calF(\calS) = \sup_{f\in\calF}\frac{1}{r}\sum_{i\in[r]}(f(s_i))^2$.
We are ready to state the technical result of this section (proof deferred to the additional materials).
\begin{theorem}\label{thm:sup_dev_rade}
Let $\calF$ be a family of functions with codomain in $[0,1]$, and let $\calS$ be a sample of $r$ random samples from a distribution $\pi$.  Denote $\hat{v}$ such that $\sup_{f\in \calF} \texttt{Var}(f)\leq \hat{v}$. For any $\delta \in (0,1)$, define 
	\begin{align}
		&	R(\calF,\calS)\leq \tilde{R} = R_r^c(\calF,\calS,\bm{\lambda}) + \sqrt{\frac{4 \mathcal{W}_{\calF}(\calS)\ln(4/\delta)}{cr}}  &\\
		& R = \tilde{R} + \frac{\ln(4/\delta)}{r}+\sqrt{\left(\frac{\ln(4/\delta)}{r}\right)^2 + \frac{2\ln(4/\delta)\tilde{R}}{r}}\nonumber & \\
		& \xi = 2R + \sqrt{\frac{2 \ln(4/\delta)\left(\hat{v}+4R\right)}{r}}+\frac{\ln(4/\delta)}{3r} \label{eq:sd_bound_rade}&
	\end{align}
	With probability at least $1-\delta$ over the choice of $\calS$ and $\bm{\lambda}$, it holds $SD(\calF,\calS)\leq \xi$.
	
\end{theorem}

%% file: trunk/approximation_algorithms.tex
\section{MANTRA: temporal Betweenness Centrality Approximation through Sampling}\label{sec:apx}
\subsection{Temporal Betweenness Estimator}\label{sec:estimators}
In this section we present one \emph{unbiased estimator} for the $(\star)$-temporal betweenness centrality, and we refer to the additional materials for the remaining estimators that have been excluded due to space constraints. The \textsc{ONBRA} (\onbra) algorithm \cite{Santoro_2022} uses an estimator defined over the sampling space $\calD_{\onbra} = \{(s,z)\in V\times V:s\neq z\}$ with uniform sampling distribution $\pi_\onbra$ over $\calD_\onbra$, and family of functions $\calF_{\onbra}$ that contains one function  $\afunc{v}{\onbra}\to [0,1]$ for each vertex $v$, defined as $\afunc{v|s,z}{\onbra} = \sigma^{(\star)}_{sz}(v)/\sigma^{(\star)}_{sz}\in [0,1]$. So far, this approach has been defined only for the shortest-temporal betweenness. In this work, we extend \onbra~to shortest-foremost and prefix foremost temporal paths.

\subsection{Sample Complexity bounds}\label{sec::ss_bound}
We present two bounds (Theorem~\ref{thm:vc_ss} and Theorem~\ref{thm::mcera_ss}) to the sufficient number of random samples to obtain an $\varepsilon$ approximation of the $(\star)$-temporal betweenness centrality.
Given a temporal graph $\calG = (V,\mathcal{E})$, with a straightforward application of Hoeffding's inequality and union bound~\cite{Mitzenmacher_2017}, it can be shown that $r= 1/(2\varepsilon^2)\log\left(2n/\delta\right)$ samples suffice to estimate the $(\star)$-temporal betweenness of every node up to an additive error $\varepsilon$ with probability $1-\delta$. To improve this bound, we define the range space associated to the $(\star)$-temporal betweeenness and its VC-dimension, and remand to~\cite{Mitzenmacher_2017,Mohri_2019,Shalev_2014} for a more complete introduction to the topic. 
Let $\calU = \dopt_\calG$, define the range space $\mathcal{R} = (\calD,\calF^+)$ where  $\calD = \calU \times [0,1]$, and $\calF^+$ is defined as follows: for a pair $(s,z)\in V\times V$ and a temporal path $tp_{sz} \in \calU$ let $f_{(v,t)}(tp_{sz}) = f((v,t)|s,z)= \mathds{1}{[(v,t)\in \textbf{{Int}}(tp_{sz})]}$ be the function that assumes value $1$ if the vertex appearance $(v,t)$ is in the temporal path between $s$ and $z$.
Moreover, define the family of functions $\calF = \{f_{(v,t)}: (v,t)\in V\times T\}$ and notice that for each $f_{(v,t)}\in \calF$ there is a range $R_{f_{(v,t)}} = \{(tp_{sz},\alpha): tp_{sz}\in \calU\land \alpha \leq f_{(v,t)}(tp_{sz})\}$. Next, define $\calF^+ =\{R_{f_{(v,t)}} :f_{(v,t)}\in \calF \}$. Now that we defined the range set for our problem, we can give an upper bound on its VC-dimension.

\begin{lemma}\label{thm:info_vc}
	The VC-dimension of the range space $\mathcal{R}^{}$ is $VC(\mathcal{R}) \leq \lfloor \log \text{VD}^{(\star)} -2\rfloor +1$.
\end{lemma}
Given the VC-dimension of the range set $\mathcal{R}$ we have:
\begin{theorem}[See~\cite{Li_2001}, Section $1$]\label{thm:vc_ss}
	Given $\varepsilon,\delta\in (0,1)$, and a universal constant $c$, let $\calS\subseteq \calD$ be a collection of elements sampled w.r.t. a probability distribution $\pi$. Then $\frac{c}{\varepsilon^2} \left[ VC(\mathcal{R})+\ln\left(\frac{1}{\delta}\right)\right]$ samples suffice to obtain $SD(\calF^+,\calS)\leq \varepsilon$ with probability $1-\delta$ over $\calS$.
\end{theorem}
To improve this bound, we make use of Lemma~\ref{lemma::diam_avg_dist} and notice that (as for the static case~\cite{Pellegrina_2021}) the $(\star)$-temporal betweenness centrality satisfies a form of negative correlation among the nodes. Moreover, the existence of a node $v$ with high $(\star)$-temporal betweenness constraints the sum of the centrality measure over the remaining nodes to be at most $\rho^{(\star)}-\gbet{v}$. In other words, this suggests that the number of nodes with high $(\star)$-temporal betweenness cannot be arbitrarily large. Furthermore, as in~\cite{Pellegrina_2021}, we assume that the maximum variance of the $(\star)$-temporal betweenness estimators $\agbet{v}$ is bounded by some estimate $\hat{v}$ rather than the worst-case upper bound of $1/4$ considered in~\cite{Borassi_2019}. This implies that the estimates $\agbet{v}$ are not bounded by the number of nodes in the temporal graph $\calG$, but are tightly constrained by the parameters $\rho^{(\star)}$ and $\hat{v}$. We are able to extend the results in~\cite{Pellegrina_2021} for the static scenario to the temporal setting for all the variants of temporal betweenness that can be computed in polynomial time and cover one of the problems left open by the authors. It follows that:
\begin{theorem}\label{thm::mcera_ss}
	Let $\calF= \{\agbet{v}, v\in V\}$ be a set of function from a domain $\calD$ to $[0,1]$. Define $\hat{v} \in (0,1/4]$ and $\rho^{(\star)}\geq 0$ such that $\max_{v\in V} \texttt{Var}(\agbet{v})\leq \hat{v}$ and $\sum_{v\in V}\gbet{v}\leq \rho^{(\star)}$.
	Fix $\varepsilon,\delta\in (0,1)$, and let $\calS$ be an $i.i.d.$ sample taken from $\calD$ of size $|\calS|\in\bigO\left(\frac{\hat{v}+\varepsilon}{\varepsilon^2}\ln\left(\frac{\rho^{(\star)}}{\delta\hat{v}}\right)\right)$.
	It holds that $SD(\calF,\calS)\leq \varepsilon$ with probability $1-\delta$ over $\calS$.
\end{theorem}
Since $\rho^{(\star)}$ correspond to the average number of internal nodes in $(\star)$-temporal paths in $\calG$, it must be that $\rho^{(\star)}\leq D^{(\star)}$. In \emph{all} the analyzed networks (see Figure~\ref{fig:diameter_and_spl} in Section~\ref{sec::experiments}) this condition holds, thus this approach will need a smaller sample size compared to the VC-Dimension based one to obtain an $\varepsilon$-approximation of the $(\star)$-temporal betweenness. 

\subsection{Fast approximation of the characteristic quantities}\label{sec::diam_apx}
According to Theorem~\ref{thm:vc_ss} and Theorem~\ref{thm::mcera_ss}, the sample size needed to achieve a desired approximation depends on the \emph{vertex diameter} and on the \emph{average temporal path length} of the temporal graph. However, under the Strong Exponential Time Hypothesis (SETH), the $(\star)$-temporal diameter (thus the average $(\star)$-temporal path length) of a temporal graph $\calG= (V,\calE)$ can not be computed in $\tilde{\bigO}(|\mathcal{E}|^{2-\varepsilon})$\footnote{With the notation $\tilde{\bigO(\cdot)}$ we ignore logarithmic factors.}~\cite{Calamai_2022}, which can be prohibitive for very large temporal graphs,
so efficient approximation algorithms for these characteristic quantities are highly desirable. Some algorithms for the diameter approximation on temporal graphs have been proposed~\cite{Crescenzi_2019,Calamai_2022}. However these techniques consider different temporal path optimality criteria~\cite{Crescenzi_2019}, or have no theoretical guarantees~\cite{Calamai_2022}. In this work we define a novel sampling-based approximation algorithm to efficiently obtain a high-quality approximation of $D^{(\star)}$ (thus, $VD^{(\star)}$) and $\rho^{(\star)}$ 
in $\tilde{O}(r\cdot |\calE|)$ where $r$ is the number of samples used by the algorithm. We provide a high level description of the sampling algorithm and we refer to the additional materials for a detailed discussion and analysis of the method. Given a temporal graph $\calG$, the sample size $r$, and the temporal path optimality $(\star)$, the algorithm performs $r$ $(\star)$-temporal BFS visits~\cite{Wu_2014} ($(\star)$-TBFS) from $r$ random nodes and keeps track of the number of reachable pairs encountered at each hop along with the greatest hop performed. Once all the $r$ visits have been completed, it computes the temporal diameter and other useful temporal measures using an approach based on the relation between the number of reachable pairs and the distance metrics~\cite{Amati_2023}. The approximation guarantees of our sampling algorithm strongly depends on ``how temporally connected'' a temporal graph is. To this end, we define the $(\star)$-temporal connectivity rate as the ratio of the number of couples that are temporally connected by a $(\star)$-temporal path and all the possible reachable couples. Formally, let $\mathds{1}[u\rightsquigarrow v]$ be the indicator function that assumes value $1$ if $u$ can reach $v$ via a $(\star)$-temporal path, then the temporal connectivity rate is defined as the ratio between the number of reachable pairs and all the possible ones in the temporal graph, i.e., $\zeta^{(\star)} = \frac{1}{n(n-1)}\sum_{u\neq v}\mathds{1}[u\rightsquigarrow v]\in [0,1]$. Intuitively, the higher the connectivity rate the higher the number of couples that are connected via at least one $(\star)$-temporal path. 
Moreover, the algorithm has the following theoretical guarantees:
\begin{theorem}\label{thm::diam_apx}
	Given a temporal graph $\calG = (V,\calE)$ and a sample of size $r = \Theta \left(\frac{\ln n}{\varepsilon^2}\right)$, the algorithm computes with probability $1-\frac{2}{n^2}$: the $(\star)$-temporal diameter $D^{(\star)}$ with absolute error bounded by $\varepsilon\over \zeta^{(\star)}$, the average $(\star)$-temporal path length $\rho^{(\star)}$ with absolute error bounded by $\frac{\varepsilon\cdot D^{(\star)}}{\zeta^{(\star)}}$, and the temporal connectivity rate with absolute error bounded by $\varepsilon$.
\end{theorem}

\subsection{The MANTRA Framework}\label{sec::mantra_framework}
In this section we introduce MANTRA\footnote{te\underline{M}por\underline{A}l betwee\underline{N}ness cen\underline{T}rality app\underline{R}oximation through s\underline{A}mpling}, our algorithmic framework for the $(\star)$-temporal betweenness centrality estimation. MANTRA incorporates the bounds in Section~\ref{sec::ss_bound} to compute an upper bound on the minimum sample size needed to approximate the SD of the $(\star)$-temporal betweenness and a state-of-the-art progressive sampling technique to speed-up the estimation process. The input parameters to MANTRA are: a temporal graph $\calG$, a temporal path optimality $(\star)\in\{\texttt{sh},\texttt{sfm},\texttt{pfm}\}$\footnote{We point out that our approach is general, and can be extended to \emph{every} definition of temporal betweenness centrality.}, a target precision $\varepsilon\in (0,1)$, a failure probability $\delta \in (0,1)$, and a number of iterations for the bootstrap phase $s'$.
The output is a vector $\calB$ of pairs $(v,\agbet{v})$ for each $v\in V$, where $\agbet{v}$ is the estimate of $\gbet{v}$ and $\calB$ is probabilistically guaranteed to be an absolute $\varepsilon$ approximation of the temporal betweenness. Formally:
\begin{theorem}\label{thm:eps_delta_guar}
	Given a target accuracy $\varepsilon\in (0,1)$ and a failure probability $\delta \in (0,1)$,
	with probability at least $1-\delta$ (over the runs of the algorithm), the output vector $\calB = \{\agbet{v}:v\in V\}$ (obtained from a set of samples $\calS$) produced by {\sc MANTRA} is such that $SD(\calB,\calS)\leq \varepsilon$.
\end{theorem}
\input{pseudocodes/adaptive_ssp_general.tex}

Algorithm~\ref{algo:prog_scheme}'s execution is divided in two phases: the bootstrap phase (lines 3-4) and the estimation phase (lines 6-15). As a first step, MANTRA, computes an upper bound $\omega$ to the number of samples needed to achieve an $\varepsilon$ approximation (line 2). The procedure runs $s'$ independent $(\star)$-TBFS visits from $s'$\footnote{In this work we use $s'= \log(1/\delta)/\varepsilon$.} random couples of nodes $(s,z)$ sampled from the population $\calD_{{\onbra}}$, estimates $\hat{v}$ and $\rho^{(\star)}$, and then plugs them in Theorem~\ref{thm::mcera_ss} to obtain $\omega$. Subsequently, it infers the first element of the sample size $\{s_i\}_{i\geq 1}$ by performing a binary search between $s'$ and $\omega$ to find the minimum $s_1$ such that Equation~\ref{eq:sd_bound_rade} (with $R$ set to $0$) is at most $\varepsilon$ and terminates the bootstrap phase. 
Such approach gives an optimistic first guess of the number of samples to process for obtaining an $\varepsilon$-approximation~\cite{Pellegrina_2021}. Subsequently it continues with the estimation phase in which, at each iteration, it increases each $s_i$ with a geometric progression~\cite{Provost_1999}, i.e., such that $s_i = 1.2\cdot s_{i-1}$. 
Next, it proceeds by drawing uniformly at random $k = s_i-s_{i-1}$ couples of nodes $(s,z)$ from $\calD_{\onbra}$ and subsequently updating the overall set of samples sampled so far (lines 7-9). Consequently, $k$ new Rademacher random vectors are added as new columns to the matrix $\bm{\lambda}$ and $k$ $(\star)$-TBFS visits are performed (line 11). Moreover, while iterating over the new sample $\mathcal{X}$ the temporal betweenness, wimpy variances and the values in $\bm{\lambda}$ are updated. After this step, the coefficients of Equation~~\ref{eq:sd_bound_rade} and the new estimate on the SD, $\xi$, are computed (lines 12-13). As a last step of the while loop, the algorithm checks whether the desired accuracy $\varepsilon$ has been achieved, i.e., whether the actual number of drawn samples is at least $\omega$ or $\xi$ is at most $\varepsilon$ (line 15). If at least one of the two conditions is met, MANTRA normalizes and outputs the current estimates $\calB$. We conclude this section with the analysis of MANTRA's running time.
\begin{theorem}\label{thm:mantra_running_time}
	Given a temporal graph $\calG = (V,E)$ and a sample of size $r$, MANTRA requires time $\tilde{\bigO}( r\cdot n\cdot |T|)$ and $\tilde{\bigO}(r\cdot |\calE|)$ to compute the shortest (foremost)-temporal and the prefix-foremost-temporal betweenness, respectively. Moreover, {\sc MANTRA} requires $\bigO(n+|\calE|)$ space.
\end{theorem}
Theorem~\ref{thm:eps_delta_guar} together with Theorem~\ref{thm:mantra_running_time} provide theoretical evidence that MANTRA computes a rigorous estimation of the $(\star)$-temporal betweenness and that \emph{scales} to the size of the input temporal graph. Moreover, it improves over the state-of-the-art approach ONBRA~\cite{Santoro_2022}. Indeed, given a sample of size $r$, ONBRA stores a $n \times r$ matrix to compute the absolute $\xi$-approximation\footnote{$\xi$ is the $SD(\calS,\calF)$ obtained using the Empirical Bernstein Bound.} using the Empirical Bernstein Bound~\cite{Maurer_2009}. Thus, ONBRA may require an arbitrary large sample size (e.g. large matrix) to achieve a target absolute approximation $\varepsilon$, making the algorithm not ideal to analyze big temporal graphs.

%% file: pseudocodes/adaptive_ssp_general.tex
\begin{algorithm}[htb!]
	\caption{MANTRA}
	\label{algo:prog_scheme} 
	\KwData{Temporal graph $\calG$, $(\star)$ temporal path optimality, precision $ \varepsilon\in (0,1)$, failure probability $\delta\in (0,1)$, bootstrap iterations $s'$, and number of Monte Carlo trials $c$.}
	\KwResult{Absolute $\varepsilon$-approximation of the $(\star)$-tbc w.p. of at least $1-\delta$. }
	
	$\calB,\mathcal{W}= [0,\dots,0 ]\in \mathbb{R}^{n}$  \tcp*{tbc and wimpy variance arrays}
	
	$i,k = 0;\xi= 1;\calS_0 = \{\emptyset\}$
	
	$\omega,\hat{v} = \texttt{DrawSufficientSampleSize}(\calG,s',\delta/2)$
	
	$\{s_i\}_{i\geq 1} = \texttt{SamplingSchedule}(\omega,\hat{v},\delta)$
	
	$\bm{\lambda} = [[\cdot] ]$ \tcp*{Empty matrix}

	
	
	\While{$true$ }{
		
		$i = i+1;k =(1.2\cdot s_{i-1})-s_{i-1}$
		
		$\mathcal{X} = \texttt{DrawSamples}(\calG,k)$\tcp{Draw $k$ samples from the sample space $\calD_{\onbra}$}
		
		$\calS_i = \calS_{i-1}\cup \mathcal{X}$
		
		$\bm{\lambda} = \texttt{Add R.R.Vector}(k,\bm{\lambda})$ \tcp{Add a Rade. rnd. column of length $c$}
		
		$\calB,\mathcal{W},\bm{\lambda}  =$\texttt{Update$(\star)$-TemporalBetweenness}$(\mathcal{X},\calB,\mathcal{W},\bm{\lambda})$	
		
%
%
%
%
%
%
		
		
		$\tilde{R},v_\calF =$\texttt{UpdateEstimates}$(\calB,\mathcal{W},\bm{\lambda},|\calS_i|,k,c)$
		
		$R^c_k = \frac{1}{c}\sum_{l = 1}^c\max_{v\in V}\left\{\tilde{R}[v,l]\right\}$
		
		$\xi =\texttt{ComputeSDBound}(R^c_k,v_\calF,\delta/2^{i},|\calS_i|)$\tcp*{Compute Eq.~\ref{eq:sd_bound_rade} in Thm.~\ref{thm:sup_dev_rade}}

		\lIf{$|\calS_i|\geq \omega $\textbf{ or }$\xi \leq \varepsilon$}{\Return $\left\{(1/|\calS_i|)\cdot \calB[u]: u \in V\right\}$}


		
	}

\end{algorithm}

%% file: trunk/experiments.tex
\section{Experimental Evaluation}\label{sec::experiments}
We compare our novel framework with ONBRA~\cite{Santoro_2022}. For the sake of fairness, we adapted the original fixed sample size algorithm to use the same progressive sampling approach of our framework. Every time an element of the sampling schedule is consumed, the algorithm computes the upper bound $\xi$ on the SD using the Empirical-Bernstein bound as in~\cite{Santoro_2022}, if $\xi$ is at most the given $\varepsilon$, it terminates, otherwise it keeps sampling. We set ONBRA's maximum number of samples to be equal to the VC-Dimension upper-bound in Section~\ref{sec::ss_bound}. We implemented all the algorithms in Julia exploiting parallel computing\footnote{Code available at: \url{https://github.com/Antonio-Cruciani/MANTRA/}}. We chose to re-implement the exact algorithms~\cite{Buss_2020} and ONBRA~\cite{Santoro_2022} because they have issues with the number of paths in the tested networks\footnote{The overflow issue appears on \emph{all} the transportation networks provided in~\cite{Kujala2018}.}, causing overflow errors (indicated by negative centralities), and with the time labeling causing an underestimation of centralities~\cite{Becker_2023}. Our implementation uses a sparse matrix representation of the $n\times |T|$ table used in~\cite{Buss_2020,Santoro_2022}, making the implemented algorithms space-efficient and usable on big temporal graphs (for which the original version of the code gives out of memory errors). We executed all the experiments on a server running Ubuntu 16.04.5 LTS with one processor Intel Xeon Gold  6248R  32 cores CPU @ 3.0GHz and 1TB RAM. For every temporal graph, we ran all the algorithms with parameter $\varepsilon\in\{0.1,0.07,0.05,0.01,0.007,0.005,0.001\}$ chosen to have a comparable magnitude to the highest temporal betweenness values in the network (see $b^{(\star)}_{\texttt{max}}$ in Table~\ref{tab::graphs}). This is a basic requirement when computing meaningful approximations\footnote{It is meaningless to compute an $\varepsilon$-approximation when the maximum centrality value is smaller than $\varepsilon$.}. Moreover, we use $\delta  =0.1$ and use $c=25$ Monte Carlo trials as suggested in~\cite{Cousins_2021,Pellegrina_2021}. Finally, each experiment has been ran $10$ times and the results have been averaged.
\input{tables/graphs}
\begin{figure}[htb!]
	\centering
		\caption{Comparison between temporal diameter and the average number of internal nodes for the Shortest (foremost) and Prefix-Foremost temporal path optimalities. The approximation has been computed (over 10 runs) using our sampling algorithm using $256$ random seed nodes.}\label{fig:diameter_and_spl}
	\begin{subfigure}{0.3\textwidth}
		\centering
		\includegraphics[scale=0.3]{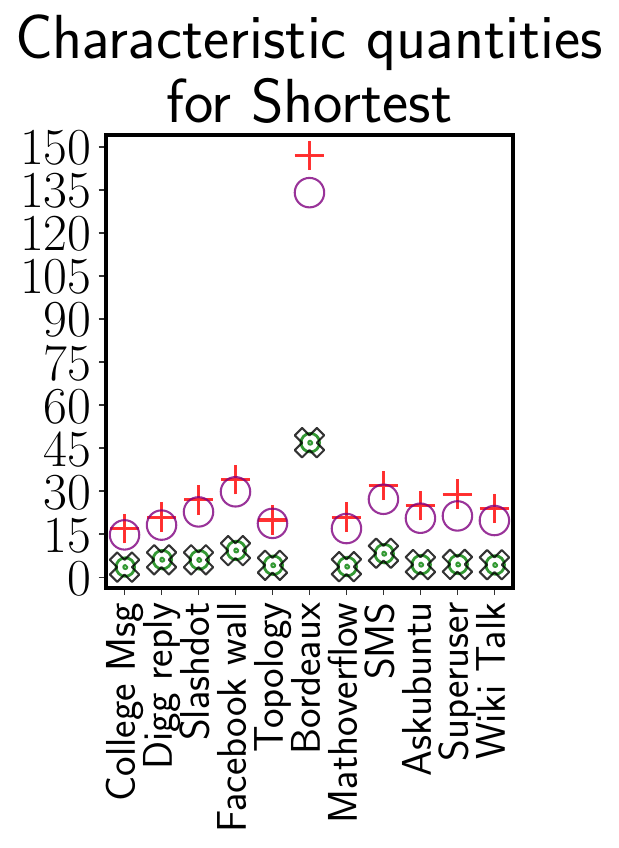}
	\end{subfigure}
	\begin{subfigure}{0.3\textwidth}
		\includegraphics[scale=0.3]{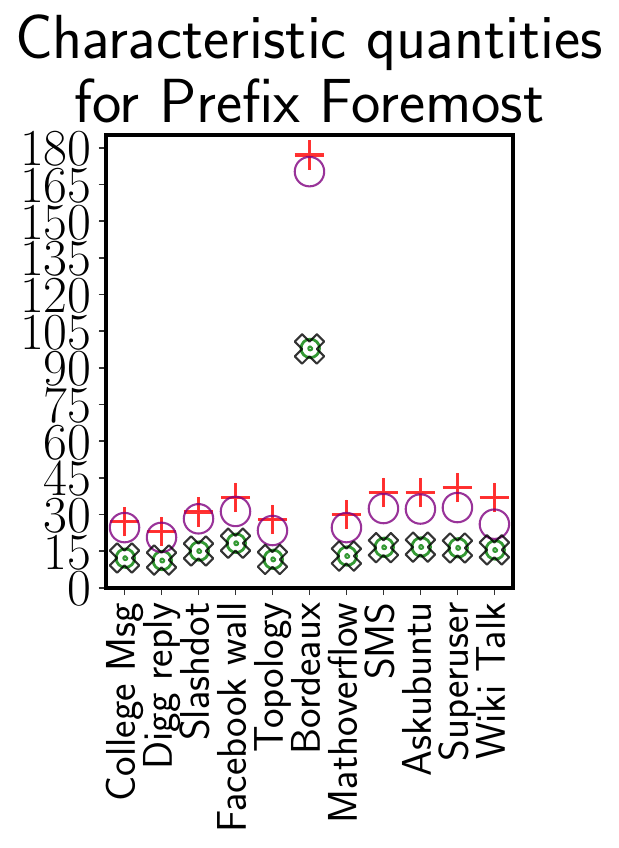}
	\end{subfigure}
	\begin{subfigure}{0.3\textwidth}
		\includegraphics[scale=0.4]{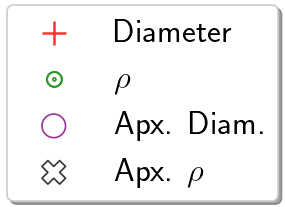}
	\end{subfigure}

\end{figure}


\subsection{Experimental Results}

\begin{figure}[htb!]

		\caption{Experimental analysis for $\varepsilon\in\{0.01,0.007,0.005,0.001\}$. Comparison between the running times  \textbf{(a)}, sample sizes \textbf{(b)}, and allocated memory \textbf{(c)} of ONBRA and MANTRA. \textbf{(d)} Supremum deviation of the absolute $\varepsilon$-approximation computed by MANTRA. The black line indicates that the two algorithms require the same amount of time/samples/memory, gray line (followed by a red mark) indicates that the algorithm required more than $1$TB of memory to run on that data set with that specific $\varepsilon$ value.}\label{fig:times_sh}

	\begin{subfigure}{0.2\textwidth}
		\centering
		\includegraphics[scale=0.13]{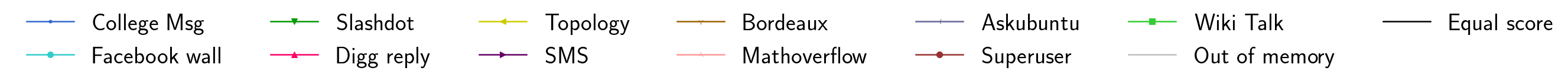}
	\end{subfigure}
	
	\begin{subfigure}{0.55\textwidth}
		\includegraphics[scale=0.22]{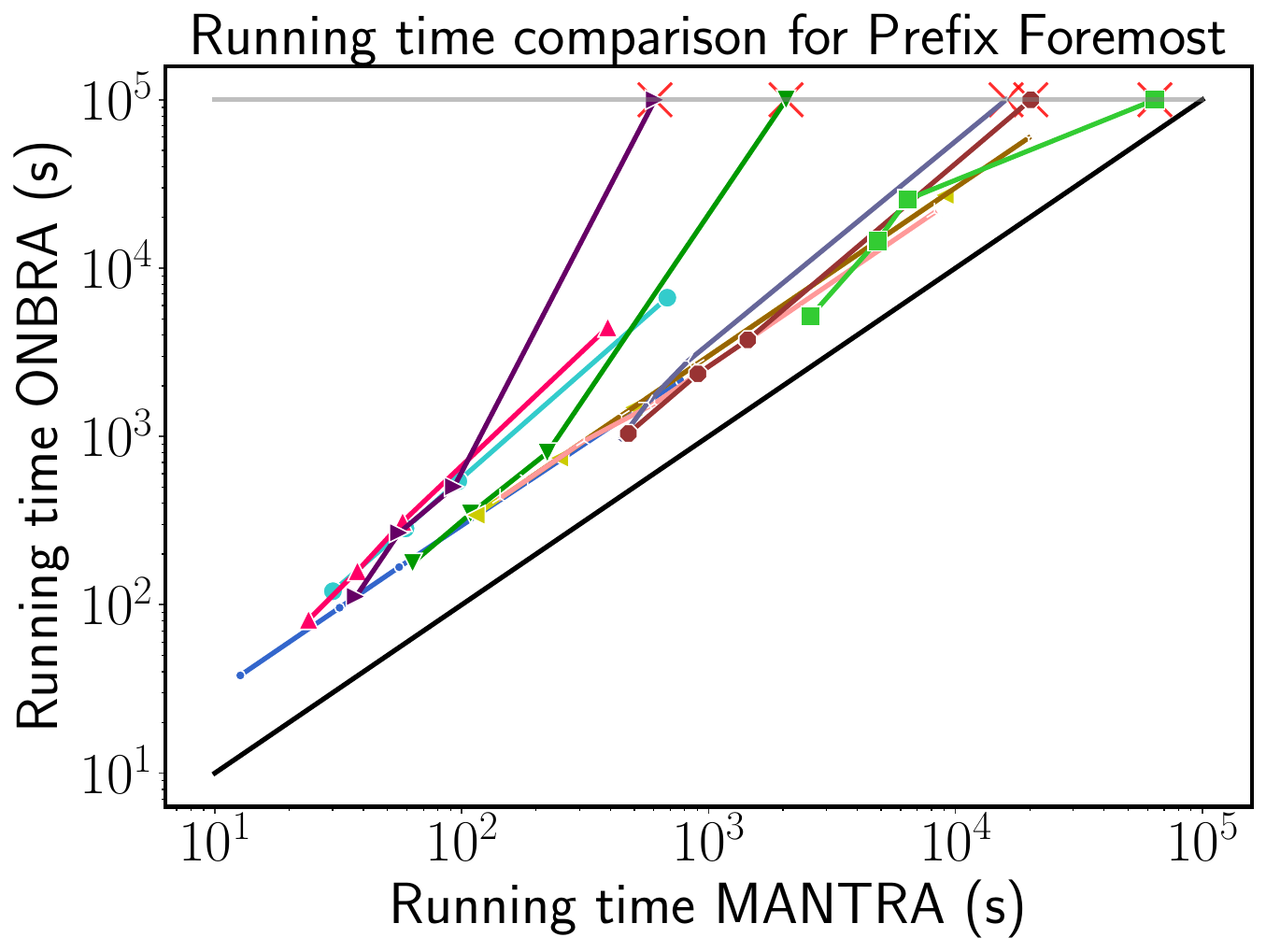}
		\caption{}\label{fig:times_pfm}
	\end{subfigure}
	\begin{subfigure}{0.45\textwidth}
		\includegraphics[scale=0.22]{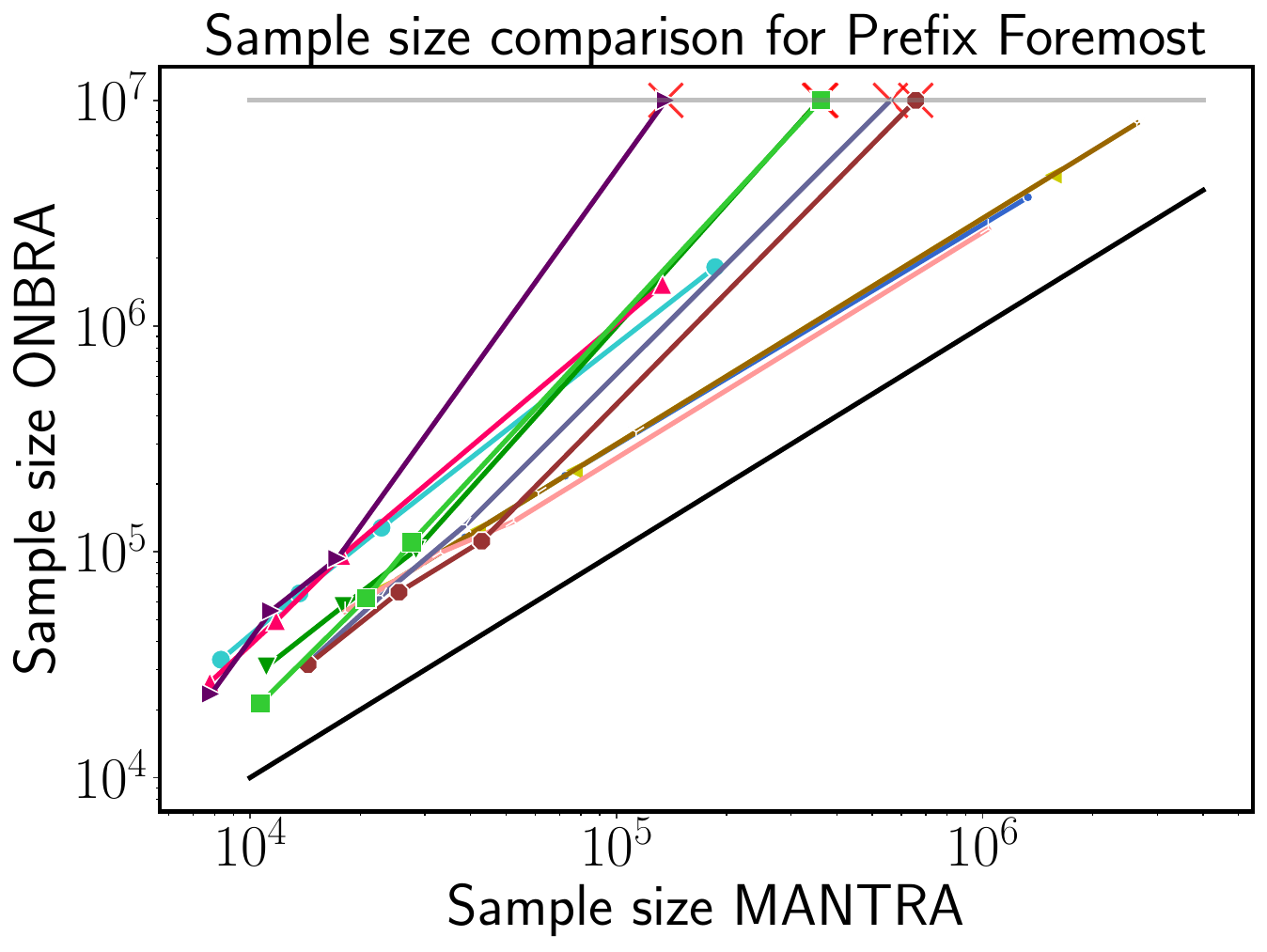}
		\caption{}\label{fig:ss_pfm}
	\end{subfigure}
	
	\begin{subfigure}{0.55\textwidth}
		\includegraphics[scale=0.22]{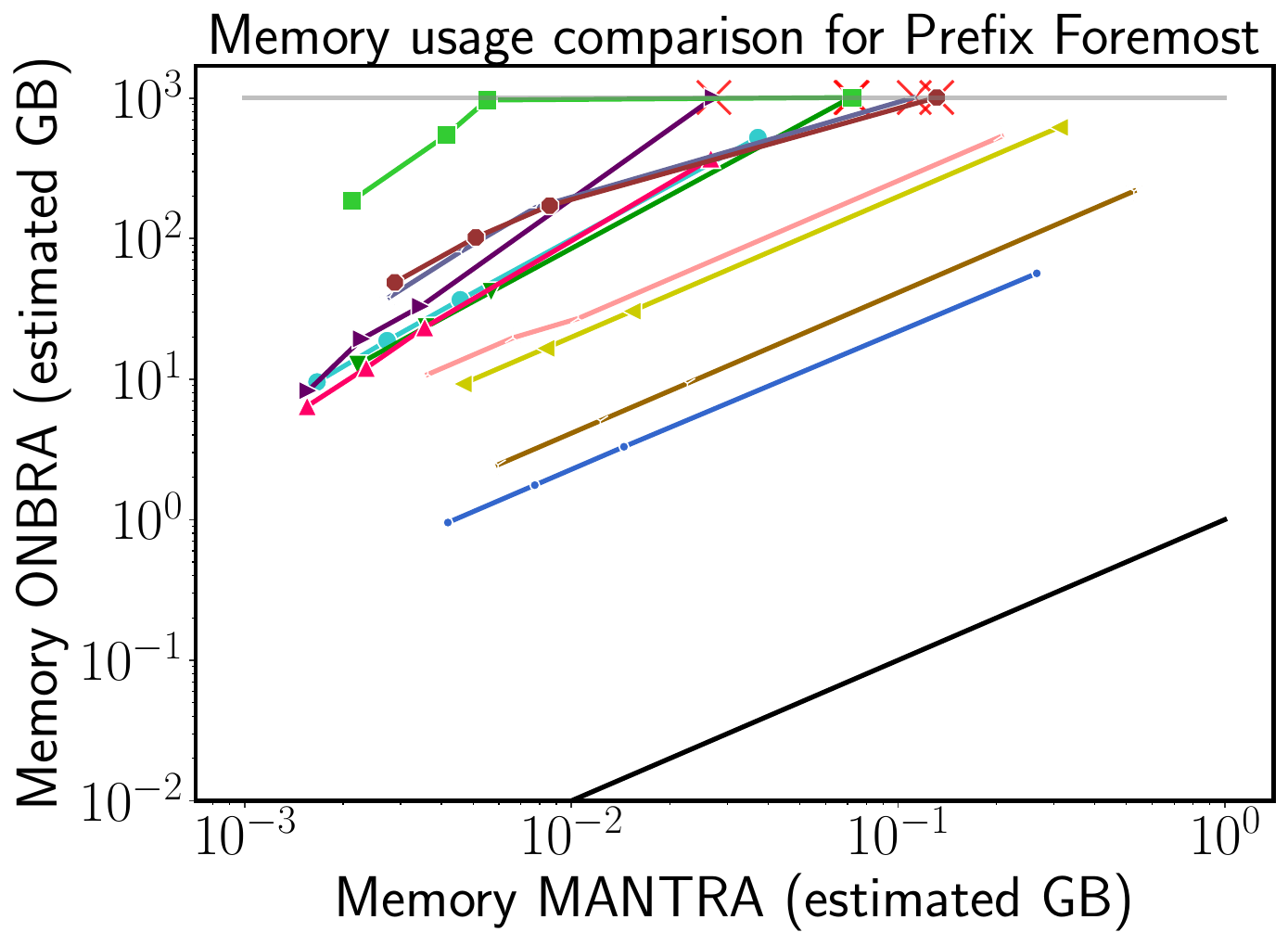}
		\caption{}\label{fig:space_pfm}
	\end{subfigure}
	\begin{subfigure}{0.45\textwidth}
		\includegraphics[scale=0.22]{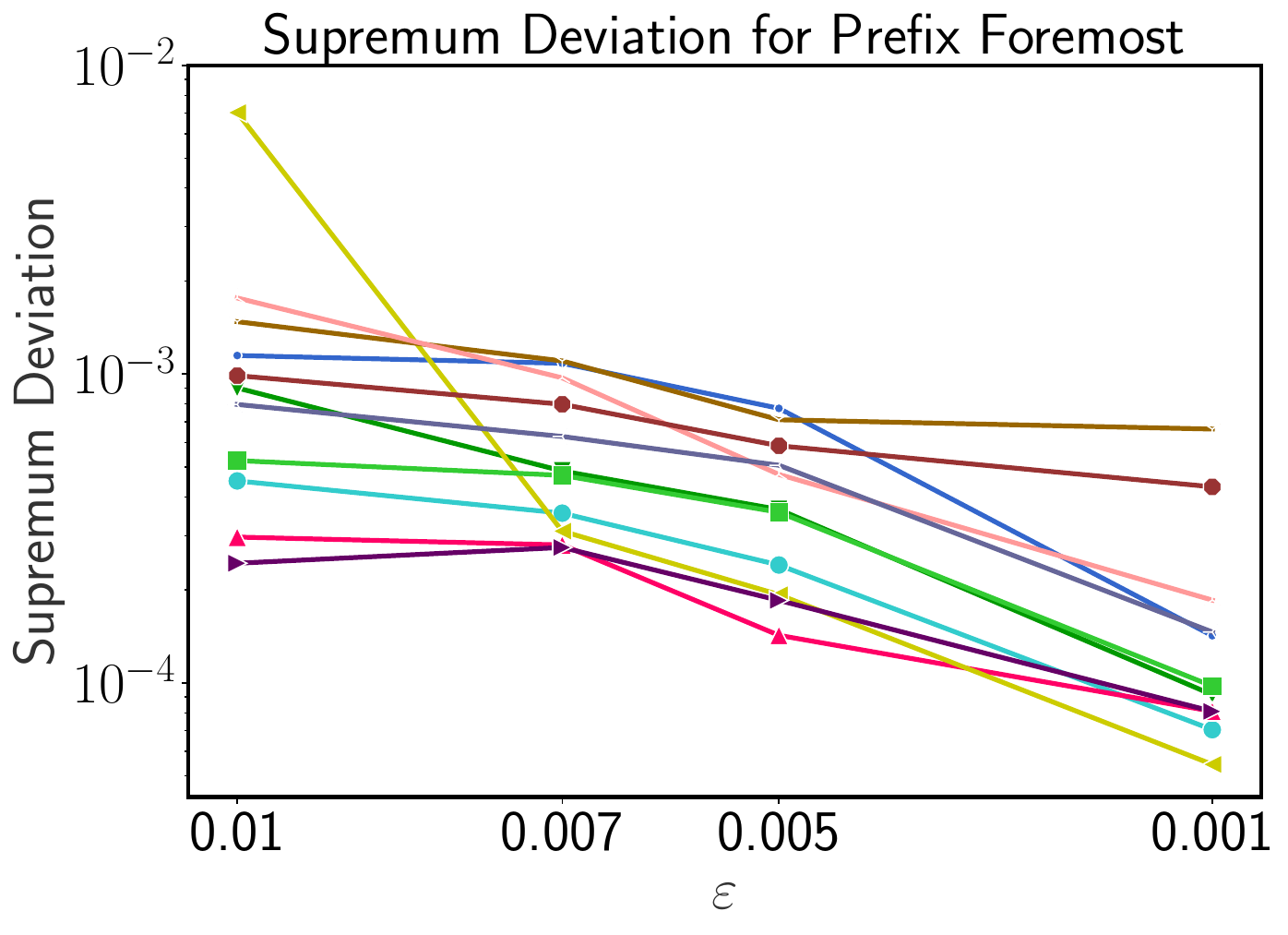}
		\caption{}\label{fig:sd_pfm}
	\end{subfigure}

\end{figure}

\paragraph{Efficiency and Scalability.}
In our first experiment, we compare the average execution times, sample sizes and allocated memory of MANTRA and ONBRA. Due to space constraints we show the results on the data sets in Table~\ref{tab::graphs} for the prefix-foremost-temporal betweenness, for a subset of $\varepsilon\in\{0.01,0.007,0.005,0.001\}$ and we refer to the additional materials for the complete battery of experiments. We chose to display the results for the \texttt{pfm} temporal path optimality because it is the one for which the analyzed graphs have the highest characteristic quantities (see Figure~\ref{fig:diameter_and_spl}). Thus, under this setting, the tested algorithms will need a bigger sample size and potentially a higher amount of memory. This somehow provides an intuitive ``upper bound'' on the algorithms performances in terms of efficiency and scalability. Moreover, the experiments for \texttt{sh} and \texttt{sfm} temporal betweenness follow similar trends of the ones displayed in the main paper. Figure~\ref{fig:times_pfm} shows the comparison of the running times (in seconds) for the \texttt{pfm} temporal betweenness. We observe that MANTRA leads the scoreboard against its competitor on \emph{all} the tested networks. Our novel framework is at least three times faster than ONBRA. Such speedup is mainly due to the smaller sample size needed to terminate. 
Furthermore, Figure~\ref{fig:ss_pfm} shows that MANTRA requires a smaller sample size (at least three times smaller) to converge. This early convergence, in practice, does not affect the approximation quality and leads to very good temporal betweenness approximations (see the next experiment). Furthermore, the number of samples needed by MANTRA varies among temporal graphs, with a strong dependence on $b^{(\star)}_{\texttt{max}}$. A potential cause of the difference in the sample sizes between the two algorithms may depend on the use of the Empirical Bernstein bound. Such bound (as the VC-Dimension one) is agnostic to any property of the analyzed temporal network, thus results in a overly conservative guarantees. This suggests that \emph{variance-adaptive} bounds are preferable to compute data-dependent approximations~\cite{Pellegrina_2021}, and that exploiting correlations among the nodes through the use of the c-MCERA leads to refined guarantees. Moreover, we point out that ONBRA does not scale well as the target absolute error $\varepsilon$ decreases. Indeed, the memory needed by ONBRA increases drastically as the target absolute error decreases (see  Figure~\ref{fig:space_pfm}) to the point of giving out of memory error for big temporal networks such as \texttt{Slashdot}, \texttt{SMS}, \texttt{Askubuntu}, \texttt{Superuser}, and \texttt{Wiki Talk}. This can lead to major issues while computing meaningful $\varepsilon$-approximations, especially under the setting in which the maximum temporal betweenness $b^{(\star)}_{\texttt{max}}$ is very small (for which we need to choose an $\varepsilon$ value of at most $b^{(\star)}_{\texttt{max}}$\footnote{We recall that $b^{(\star)}_{\texttt{max}}$ can be efficiently approximated in the bootstrap phase of our framework.}). Unfortunately, this is not an uncommon feature of real-world temporal networks. Indeed, as shown by $\zeta$ and $b^{(\star)}_{\texttt{max}}$ in Table~\ref{tab::graphs} they tend to be very sparse. This experiment, suggests that MANTRA is preferable for analyzing big temporal networks up to an arbitrary small absolute error $\varepsilon$. 

\paragraph{Comparison with the exact algorithms scores and running times.}

\begin{figure}[htb!]
	\caption{\textbf{(a)} Relation between the running time and the sample size of MANTRA for the shortest temporal betweenness with  $\varepsilon$ as in Figure~\ref{fig:times_sh}. \textbf{(b)} Comparison between MANTRA and the exact algorithm running times for the shortest temporal betweenness on the biggest temporal networks.}\label{fig:times_ex_sh}
	
	\begin{subfigure}{0.2\textwidth}
		\centering
		\includegraphics[scale=0.15]{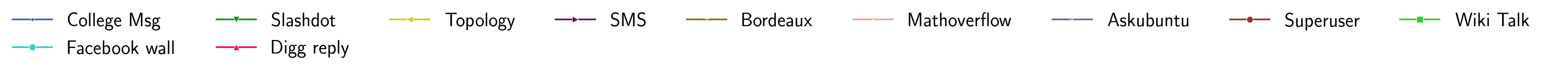}
	\end{subfigure}
	
	\begin{subfigure}{0.55\textwidth}
		\includegraphics[scale=0.22]{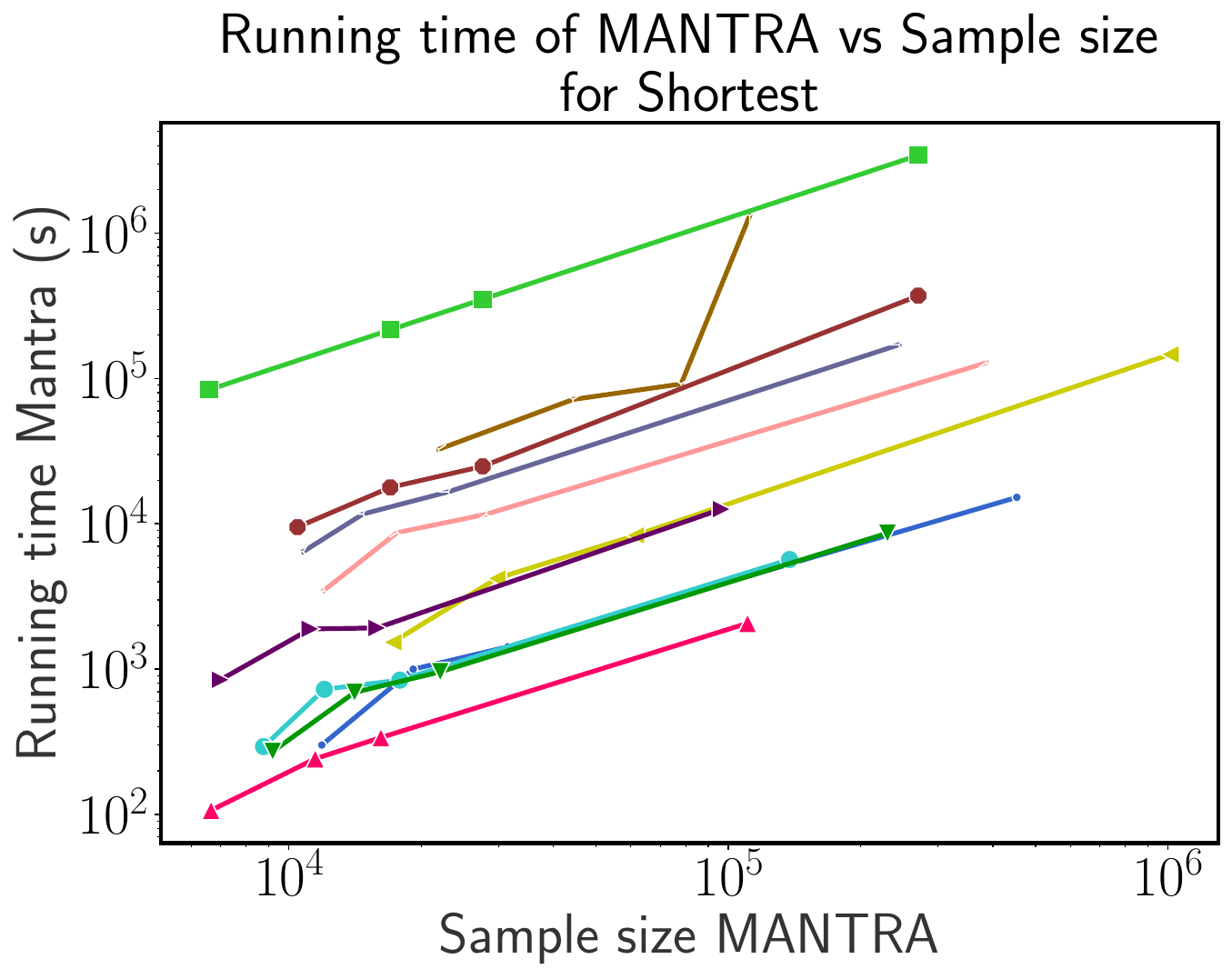}
		\caption{}\label{fig:time_vs_ss}
	\end{subfigure}
	\begin{subfigure}{0.45\textwidth}
		\includegraphics[scale=0.22]{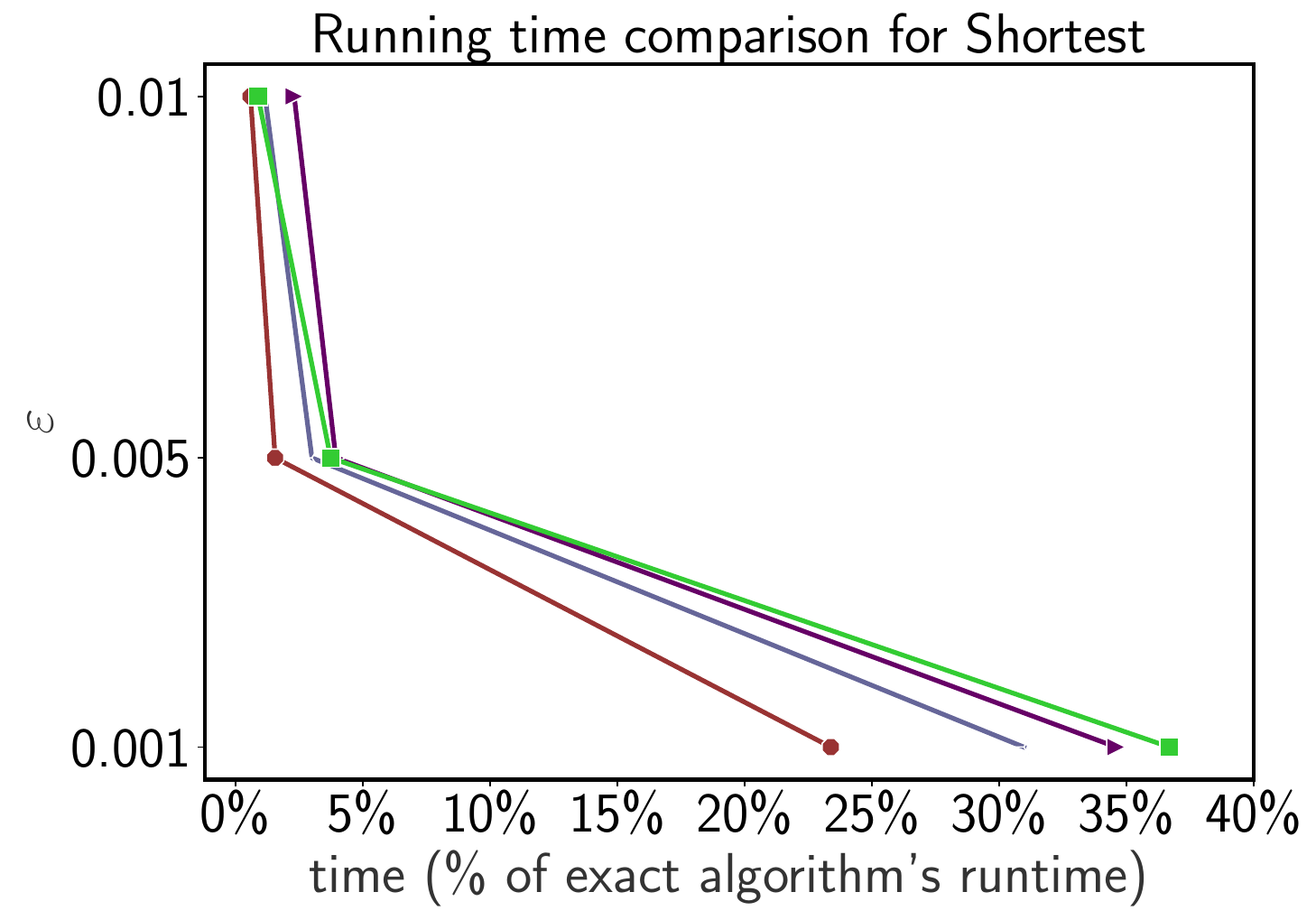}
		\caption{}\label{fig:time_ex_mantra}
	\end{subfigure}	
\end{figure}

As a first step in our second experiment, we investigate the accuracy of the approximations provided by MANTRA by computing the exact temporal betweenness centrality of all the nodes of the temporal network in Table~\ref{tab::graphs} and measuring the SD over all the ten runs. Figure~\ref{fig:sd_pfm} supports our theoretical results, as we always get a SD of at most the given $\varepsilon$. 
Moreover, we point out that the exact algorithms for the shortest (foremost) temporal betweenness required a time that spanned from several hours (e.g. for \texttt{SMS}) to days (for \texttt{Askubuntu}, and \texttt{Superuser} $\approx$ a week) and weeks (for \texttt{Wiki Talk} $\approx$ a month). Instead, MANTRA completes the approximation in reasonable time. Figure~\ref{fig:time_vs_ss} shows the relation between the sample size and the running time of our framework. While, Figure~\ref{fig:time_ex_mantra} shows the amount of time needed by MANTRA to provide the absolute $\varepsilon$-approximation in terms of percentage of exact algorithm's running time. We display the running times on the biggest temporal graphs for the \texttt{sh} temporal betweenness because is one of the ``critical'' temporal path optimalities that requires longer times to be computed (see Theorem~\ref{thm:mantra_running_time}). We can conclude that our framework is well suited to quickly compute \emph{effective} approximations of the temporal betweenness on very large temporal networks.

%% file: tables/graphs.tex
\begin{table}[htb!]

	\centering
		\caption{The data sets used in our evaluation, where $\zeta$ indicates the exact temporal connectivity rate, $b^{(\star)}_{\texttt{max}}$ the maximum $(\star)$-temporal betweenness centrality (type D stands for directed and U for undirected). $\bullet$ indicates that we need to use \texttt{BigInt} data type instead of \texttt{Unsigned Int128} to count the number of shortest (foremost)-temporal paths to avoid overflows.}\label{tab::graphs}
	\begin{tabular}{llllllllcc}
		\toprule
		\textbf{Data set}    & \multicolumn{1}{l}{\bm{$n$}} & \multicolumn{1}{l}{\bm{$|\calE|$}} & \multicolumn{1}{l}{\bm{$|T|$}} & \multicolumn{1}{l}{\bm{$\zeta$}} &\multicolumn{1}{c}{$\bm{b}^{(\texttt{pfm})}_{\texttt{max}}$}&\multicolumn{1}{c}{$\bm{b}^{(\texttt{sh})}_{\texttt{max}}$}&\multicolumn{1}{c}{$\bm{b}^{(\texttt{sfm})}_{\texttt{max}}$}& \multicolumn{1}{c}{\textbf{Type}}& \multicolumn{1}{c}{\textbf{Source}}  \\ \hline
		\texttt{College msg}      & 1899                               & 59798                           & 58911                          & 0.5                                & 0.0718 & 0.0319 & 0.0365& D                              &\cite{SnapnetsWebSite}  \\
		\texttt{Digg reply}       & 30360                              & 86203                           & 82641                          & 0.02                               & 0.0019 & 0.0015 & 0.0016&D                               & \cite{TemporalNetworkRepositoryWebsite} \\
		\texttt{Slashdot}   & 51083                              & 139789                          & 89862                          & 0.07                              & 0.0128 & 0.0074 & 0.0085 &D                                & \cite{TemporalNetworkRepositoryWebsite}\\
		\texttt{Facebook Wall}    & 35817                              & 198028                          & 194904                         & 0.04                              & 0.0034 & 0.0024 & 0.0028&D                               & \cite{TemporalNetworkRepositoryWebsite} \\
		\texttt{Topology}          & 16564                              & 198038                          & 32823                          & 0.53                               & 0.0921 & 0.0654 & 0.0681  &U                                 &\cite{KonectWebSite} \\
		\texttt{Bordeaux}$^\bullet$          & 3435                               & 236075                          & 60582                          & 0.84                               & 0.1210 & 0.1383 & 0.1269 &D                              & \cite{Kujala2018}  \\
		\texttt{Mathoverflow}      & 24759                              & 390414                          & 389952                         &                   0.33                 & 0.0522 & 0.0282 & 0.0287&D                          &  \cite{SnapnetsWebSite}     \\
		\texttt{SMS}               & 44090                              & 544607                          & 467838                         & 0.008                             & 0.0019 & 0.0010 & 0.0012 &D                              & \cite{SnapnetsWebSite}  \\
		\texttt{Askubuntu}         & 157222                             & 726639                          & 724715                         & 0.169                                  & 0.0214 & 0.0156 & 0.0154 &D                         &  \cite{SnapnetsWebSite}      \\
		\texttt{Super user}        & 192409                             & 1108716                         & 1105102                        &             0.21                        & 0.0261 & 0.0165 & 0.0182&D                          &    \cite{SnapnetsWebSite}   \\
		\texttt{Wiki Talk}        & 1094018                             & 6092445                         & 5799206                        &                       0.069              & 0.0089&0.0155  &0.0153 &D                          &    \cite{KonectWebSite}   \\
		
		\bottomrule                             
	\end{tabular}

\end{table}

%% file: trunk/conclusions.tex
\section{Conclusions}
We proposed MANTRA, a novel framework for approximating the temporal betweenness centrality on large temporal networks. MANTRA relies on the state-of-the-art bounds on supremum deviation of functions based on the c-MCERA to provide a probabilistically guaranteed absolute $\varepsilon$-approximation of such centrality measure. Our framework includes a fast sampling algorithm to approximate the temporal diameter, average path length and connectivity rate up to a small error with high probability. Such approach is general and can be adapted to approximate several version of these quantities based on different temporal path optimalities (e.g.~\cite{Crescenzi_2019,Calamai_2022}). Our experimental results (summarized in Section~\ref{sec::experiments}) depict the performances of our framework versus the state-of-the-art algorithm for the temporal betweenness approximation. MANTRA consistently over-performs its competitor in terms of running time, sample size, and allocated memory. As indicated in Figure~\ref{fig:space_pfm}, our framework is the only available option to obtain meaningful temporal betweenness centrality approximations when we do not have access to servers with a large amount of memory. In the spirit of reproducibility, we developed an open source framework in Julia that allows any user with an \emph{average laptop} to approximate the temporal betweenness centrality on any kind of graph. Some promising future directions are to use MANTRA to find communities in temporal graphs and to extend our approach to other temporal path based centrality measures.

%% file: trunk/appendix.tex
\section{Extension of the $(\star)$-Temporal Betweenness Centrality to the edges.}\label{apx::temporal_edge_betweenness}
\begin{figure}[htb!]
	\centering
	\includegraphics[width=6cm]{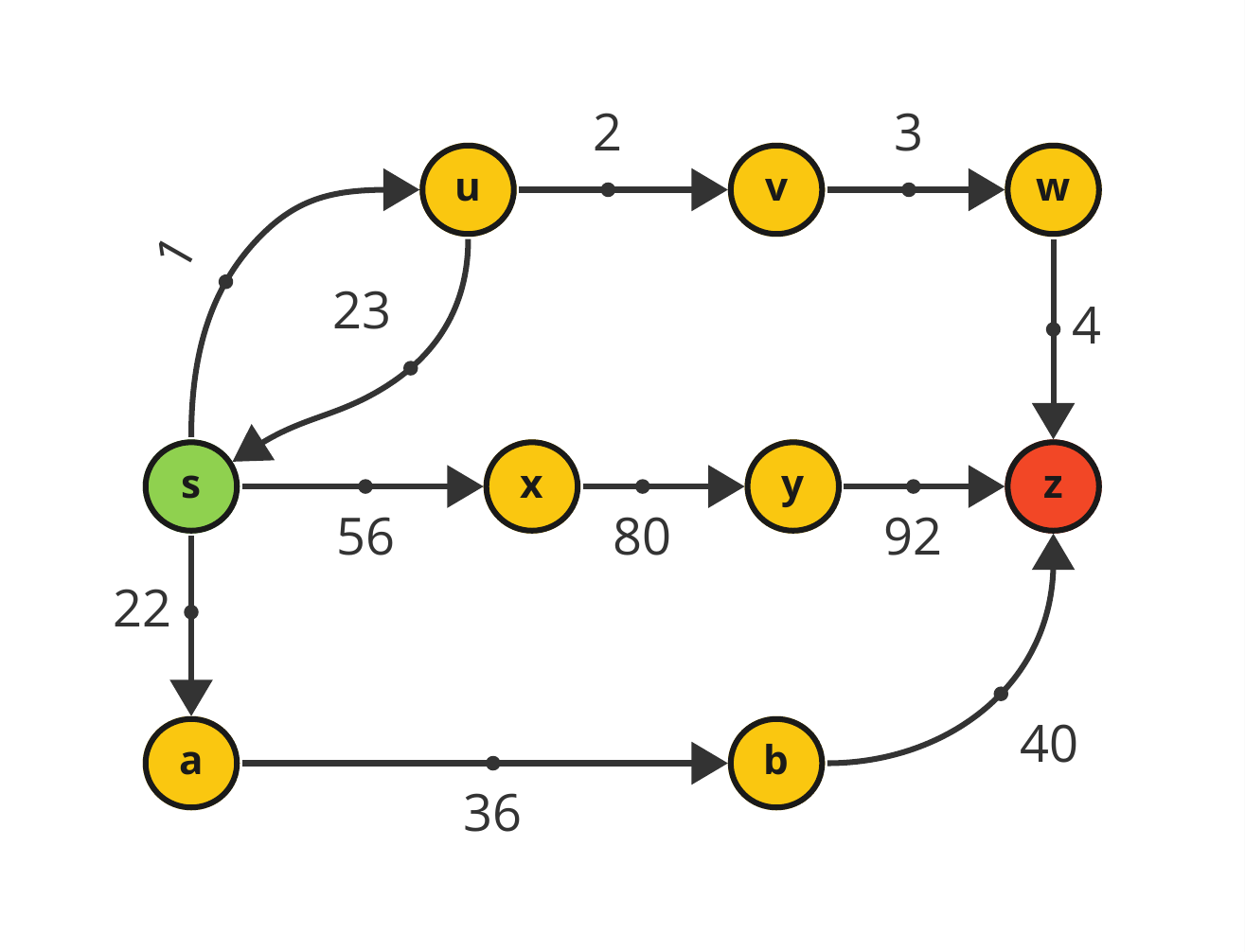}
	\caption{Example of the $(\star)$-temporal paths described in Definition~\ref{def::tps}. \textbf{Shortest}: $(s\xrightarrow{56}x\xrightarrow{80}y\xrightarrow{92}z),(s\xrightarrow{22}a\xrightarrow{36}b\xrightarrow{40}z)$, \textbf{Shortest-Foremost}: $(s\xrightarrow{22}a\xrightarrow{36}b\xrightarrow{40}z)$, and \textbf{Prefix-Foremost}:$(s\xrightarrow{1}u\xrightarrow{2}v\xrightarrow{3}w\xrightarrow{4}z)$.}\label{fig:temporal_paths}
\end{figure}
Figure~\ref{fig:temporal_paths} shows an example of the temporal paths optimalities considered in this paper. Moreover, we define the \emph{underlying graph} as the \emph{static} graph $G= (V,E)$ obtained by removing all the time instants and multi-arcs from the set $\calE$.
We extend the concept of temporal betweenness to the temporal and underlying edge of a given temporal graph $\calG= (V,\calE)$ as follows
\begin{definition}[Temporal Edge Betweenness]
	The temporal betweenness of any edge $e=(u,v)$ in the \emph{underlying graph} of a temporal graph $\mathcal{G}$ is defined as:
	\begin{align}\label{eq:temporal_edge_bc}
		\gbet{e} = \frac{1}{n(n-1)}\sum_{s\neq z} \frac{\sigma_{sz}^{(\star)}(e)}{\sigma_{sz}^{(\star)}}
	\end{align}
\end{definition}
We define the temporal betweenness of a temporal edge as follows:
\begin{align}
	\gbet{e,t} =  \frac{1}{n(n-1)}\sum_{s\neq z} \frac{\sigma_{sz}^{(\star)}(e,t)}{\sigma_{sz}^{(\star)}} 
\end{align}
Where $\sigma_{sz}^{(\star)}(e,t)$ is the number of \optpaths from $s$ to $z$ passing through the underlying edge\footnote{Equivalently, let $e= (u,v)$, then $\sigma_{sz}^{(\star)}(e,t)$ is the number of \optpaths from $s$ to $z$ passing through the temporal edge $(u,v,t)$. } $e$ at time $t$.
Now we define the temporal edge betweenness of an underlying edge $e$ as the sum of the temporal edge betweenness values of its appearances at time $(e,t)$
\begin{lemma}\label{lemma:temporal_edge}
	For any underlying edge $e\in E$ it holds:
	\begin{align} 
		\gbet{e}= \frac{1}{n(n-1)}\sum_{t=0}^T\gbet{e,t}
	\end{align}
\end{lemma} 
\begin{proof}
	\begin{align*}
		&	\gbet{e} =\frac{1}{n(n-1)} \sum_{s\neq z} \frac{\sigma_{sz}^{(\star)}(e)}{\sigma_{sz}^{(\star)}} =\frac{1}{n(n-1)} \sum_{s\neq z} \sum_{t=0}^T \frac{\sigma_{sz}^{(\star)}(e,t)}{\sigma_{sz}^{(\star)}} = \frac{1}{n(n-1)} \sum_{t=0}^T\gbet{e,t}&
	\end{align*}
\end{proof} 
Analogously to Lemma~\ref{lemma::diam_avg_dist}, we can show that the sum of the $(\star)$-temporal betweenness centrality of the underlying edges is equal to the average number of edges in a $(\star)$-temporal path. Given a temporal graph $\calG= (V,\calE)$ define
\begin{align*}
	\Psi^{(\star)} = \frac{1}{n(n-1)}\sum_{s,z\in V } \sum_{e\in E}\mathds{1}{[e\in tp_{sz}]}
\end{align*}
Where $\mathds{1}{[e\in tp_{sz}]}$ assumes value $1$ if and only if the underlying edge $e\in E$ appears in the temporal path $tp_{sz}$. Then the following lemma holds:
\begin{lemma}\label{lemma:avg_dist_edges}
	$\sum_{e\in E}\gbet{e} = \Psi^{(\star)}$
\end{lemma}
\begin{proof}
	Equation~\ref{eq:temporal_edge_bc} can be rewritten as 
	$$\gbet{e} =\frac{1}{n(n-1)} \sum_{s,z\in V}\sum_{tp\in \sopt{sz}}\frac{\mathds{1}{[e\in \textbf{\textbf{Int}}(tp)]}}{\sigma^{(\star)}_{sz}}$$
	Summing over the underlying edges $e\in E$ we obtain:
	\begin{align*}
		&\sum_{v\in V}\gbet{e} =\frac{1}{n(n-1)} \sum_{s,z\in V}\sum_{tp\in \sopt{sz}}\frac{1}{\sigma^{(\star)}_{sz}}\sum_{e\in E}{\mathds{1}{[e\in \textbf{\textbf{Int}}(tp)]}} &\\& \frac{1}{n(n-1)} \sum_{s,z\in V}\frac{\sigma^{(\star)}_{sz}}{\sigma^{(\star)}_{sz}}\sum_{e\in E}{\mathds{1}{[e\in \textbf{\textbf{Int}}(tp_{sz})]}}=\frac{1}{n(n-1)} \sum_{s,z\in V} \sum_{e\in E}{\mathds{1}{[e\in \textbf{\textbf{Int}}(tp_{sz})]}} = \Psi^{(\star)}&
	\end{align*}
	
\end{proof}

Given Lemma~\ref{lemma:temporal_edge}, Lemma~\ref{lemma:avg_dist_edges}, and the temporal accumulation results in Bu{\ss} et al. \cite{Buss_2020}, we can adapt all the algorithms\footnote{Exact and approximate ones.} for computing the $(\star)$-Temporal Vertex Betweenness centrality to compute the  $(\star)$-Temporal Edge Betweenness of the underlying graph. Edge-temporal betweenness centrality can be used to develop fast temporal-community detection algorithms by using an approach similar to the well know Girwan-Newman  algorithm~\cite{Girvan_2002}. Our approximation algorithms can be used to efficiently partition temporal graphs by removing the top-k underlying/temporal edges with highest temporal-betweenness scores and find communities.

\section{Missing proofs}\label{apx:missing_proofs}
\paragraph{\textbf{Proof of Lemma~\ref{lemma::diam_avg_dist}}.}
\begin{proof}
	We can write the $(\star)$-temporal betweenness as
	$$\gbet{v} =\frac{1}{n(n-1)}\sum_{\substack{s\neq v\neq z }}{\frac{\sigma_{sz}^{(\star)}(v)}{\sigma^{(\star)}_{sz}}}=\frac{1}{n(n-1)} \sum_{s,z\in V}\sum_{tp\in \sopt{sz}}\frac{\mathds{1}{[v\in \textbf{\textbf{Int}}(tp)]}}{\sigma^{(\star)}_{sz}}$$
	summing over all the nodes, we obtain 
	\begin{align*}
		&\sum_{v\in V}\gbet{v} =\frac{1}{n(n-1)} \sum_{s,z\in V}\sum_{tp\in \sopt{sz}}\frac{1}{\sigma^{(\star)}_{sz}}\sum_{v\in V}{\mathds{1}{[v\in \textbf{\textbf{Int}}(tp)]}} &\\&= \frac{1}{n(n-1)} \sum_{s,z\in V}\frac{\sigma^{(\star)}_{sz}}{\sigma^{(\star)}_{sz}}\sum_{v\in V}{\mathds{1}{[v\in \textbf{\textbf{Int}}(tp_{sz})]}}=\frac{1}{n(n-1)} \sum_{s,z\in V}\left|\textbf{Int}(tp_{sz})\right| = \rho^{(\star)}&
	\end{align*}
\end{proof}
\begin{lemma}\label{lemma:onbra}
	The function computed by \onbra~is an unbiased estimator of the $(\star)$-temporal betweenness centrality.
\end{lemma}
\begin{proof}
	\begin{align*}
		&\Expec{\afunc{v|s,z}{\onbra}} =\sum_{\substack{s,z\in V\\ s\neq z\neq v}} \Prob{(s,z)}\cdot \afunc{v|s,z}{\onbra} 
		=\sum_{\substack{s,z\in V\\ s\neq z}}\frac{1}{n(n-1)}  \frac{\sigma_{sz}^{(\star)}(v)}{\sigma_{sz}^{(\star)}}&
	\end{align*}
\end{proof}

\paragraph{\textbf{Proof of Lemma~\ref{thm:info_vc}}}
\begin{proof}
	Let $VC(\mathcal{R}) = d$, where $d\in \mathbb{N}$. Then, there is $S\subset \calD$ such that $|S|=d$ and $S$ is shattered by $\calF^{+}$. For each temporal path $tp_{sz}\in \calU$, there is at most one pair $(tp_{sz},\alpha)$ in $S$ for some $\alpha\in (0,1]$ and there is no pair of the form $(tp_{sz},0)$. By definition of shattering, each $(tp_{sz},\alpha)\in S$ appears in $2^{d-1}$ different ranges in $\calF^{+}$. Moreover, each pair $(tp_{sz},\alpha)$ is in at most $|tp_{sz}|-2$ ranges in $\calF^+$, that is because $(tp_{sz},\alpha)\notin R_{f(v,t)}$ either when $\alpha > f_{(v,t)}(tp_{sz})$ or $(v,t)\notin tp_{sz}$. Observe that $|tp_{sz}|-2\leq \text{VD}^{(\star)}-2$, gives $2^d \leq |tp_{sz}|-2\leq \text{VD}^{(\star)}-2 $.
	Thus,
	$d-1 \leq \log (\text{VD}^{(\star)}-2)$ since $d\in\mathbb{N}$, we have
	$d\leq \lfloor \log \text{VD}^{(\star)} -2\rfloor +1 \leq \log (\text{VD}^{(\star)}-2)+1$.
	Finally,
	$ VC(\mathcal{R}) = d\leq  \lfloor \log \text{VD}^{(\star)} -2\rfloor +1$.
\end{proof}
We show how to estimate $\hat{v} = \sup_{f\in \calF}\Var{f}$ using the empirical wimpy variance $\mathcal{W}_\calF(\calS)$. Proposition~\ref{lemma:self_bounding_wimpy} is an adaption of Proposition 4.3~\cite{Pellegrina_2021} when we are using only a unique family of function $\calF$ rather than a partition. For completeness we provide the adapted proposition and the proof. 
\begin{proposition}[\cite{Pellegrina_2021}]\label{lemma:self_bounding_wimpy}
	Let $\calF= \{\agbet{v}, v\in V\}$ be a set of function from a domain $\calD$ to $[0,1]$. And let $\calS\subseteq \calD$ be a sample of size $r$. Then, with probability at least $1-\delta$ it holds
	\begin{align}
		&\sup_{f\in \calF}\Var{f}\leq \mathcal{W}_\calF(\calS)+\frac{\ln(1/\delta)}{r} + \sqrt{\left(\frac{\ln(1/\delta)}{r}\right)^2 + \frac{\mathcal{W}_\calF(\calS)+\ln(1/\delta)}{r}}&
	\end{align}
\end{proposition}
\begin{proof}
	By definition, 
		\begin{align}
	&\sup_{f\in \calF}\Var{f} = \sup_{f\in \calF}\left\{\Expec{f^2}-\Expec{f}^2\right\}\leq \sup_{f\in \calF}{\Expec{f^2}}&
		\end{align}
		Thus we focus on bounding $\sup_{f\in \calF}{\Expec{f^2}}$. By a straightforward application of Theorem 7.5.8 in~\cite{Pellegrina_2021_b} we have the claim of the lemma. 
\end{proof}

\paragraph{\textbf{Proof of Theorem~\ref{thm::mcera_ss}}.} 
The proof of this theorem follows the one in~\cite{Pellegrina_2021} by considering the properties of the $(\star)$-temporal betweenness. For completeness, we show the adapted proof.
\begin{proof}
Given a sample $S$ of size $s$, let $E$ and $E_v$ be the following events:
\begin{align*}
	& E = ``\exists v\in V : |\gbet{v}-\agbet{v}|>\varepsilon\text{''}&\\&
 E_v = `` |\gbet{v}-\agbet{v}|>\varepsilon\text{''}&
\end{align*}
Applying the union bound we have that $\Prob{E} \leq \sum_{v\in V}\Prob{E_v}$. Define the functions $g(x) = x(1-x)$ and $h(x) = (1+x)\ln (1+x)-x$ for $x\geq 0$, and let $\hat{x}_1,\hat{x}_2$, $\hat{x}$ as \begin{align*}
	&\hat{x}_1 = \inf\left\{x:\frac{1}{2}-\sqrt{\frac{\varepsilon}{2}-\frac{\varepsilon^2}{9}}\leq x\leq \frac{1}{2},g(x)h\left(\frac{\varepsilon}{g(x)}\right)\left( 2\varepsilon^2\right)\right\}&\\&\hat{x}_2 = \frac{1}{2}-\sqrt{\frac{1}{4}-\hat{v}}\quad\quad\text{and},\quad\quad \hat{x} = \min\{\hat{x}_1,\hat{x}_2\}&
	\end{align*}
Moreover, using Hoeffding's and Bennet's bounds~\cite{Boucheron_2013}, Bathia and Davis bound on variance~\cite{Bhatia_2000} and from the fact that 

$\max_{v\in V} \texttt{Var}(f(v))\leq \hat{v}$, it holds, for all $v\in V$
\begin{align*}
&\Prob{E_v}\leq  2\min\left\{\exp\left(-2s\varepsilon^2\right),\exp\left(-s\gamma(\hat{v},\gbet{v},\varepsilon) \right)\right\} \doteq \eta(x) &	
\end{align*}
where $\gamma(\hat{v},\gbet{v},\varepsilon) = \min\{\hat{v},g(\gbet{v})\}h\left(\frac{\varepsilon}{\min\{\hat{v},g(\gbet{v})\}}\right)$. 
Now we can write
\begin{align}\label{eq::bound_silv}
	\Prob{E} \leq \sum_{v\in V}E_v = \sum_{v\in V}\Phi(\gbet{v})
\end{align}
Observe that the values of $\gbet{v}$ are not known a priori, thus it is not possible to compute the r.h.s. of Equation~\ref{eq::bound_silv}. However, we can obtain a sharp upper bound by using the constraints on the possible values of $\gbet{v}$ imposed by $\hat{v}$ and $\rho^{(\star)}$. As in~\cite{Pellegrina_2021} we define an appropriate optimization problem w.r.t. the unknown values of $\gbet{v}$. Let $k_x$ be the number of nodes that we assume have $\gbet{v} = x$, define the optimization problem over the variables $k_x$ as follows:

\begin{align*}
	\begin{array}{ll@{}ll}
		\text{max}  & \displaystyle\sum\limits_{\substack{x\in (0,1)\\ k_x >0}}k_x\Phi(x) &\\
		\text{subject to}& \displaystyle\sum\limits_{\substack{x\in (0,1)\\ k_x >0}}   xk_x \leq \rho^{(\star)},  &\\
		&              0\leq k_x \leq \frac{\rho^{(\star)}}{x},k_x\in\mathbb{N}                                  &
	\end{array}
\end{align*}
Observe that the first constraint follows from Lemma~\ref{lemma::diam_avg_dist}, and the second one directly by the definition of $\rho^{(\star)}$ itself. The values of the objective function of the optimal solution of the optimization problem give and upper bound to Equation~\ref{eq::bound_silv}. As for the static case~\cite{Pellegrina_2021}, the optimization problem is a formulation of the Bounded Knapsack Problem~\cite{Martello_1990} over the variables $k_x$ in which items with label $x$ are selected $k_x$ times with unitary profit $\Phi(x)$ and weight $x$. Moreover, we consider the upper bound to the optimal solution given by the continuous relaxation, in which $k_x\in\mathbb{R}$, of the optimization problem (see Chapter 3 of~\cite{Martello_1990}). Four our purpose. its enough to fully select the item with higher profit-weight ratio to fill the entire knapsack. Let $\overline{x} = \argmax_{x\in (0,1)} \{\Phi(x)/x\} $, the optimal solution of the continuous relaxation is $k_x = \rho^{(\star)}/\overline{x}, k_x = 0$, for all $x\neq \overline{x}$, while the optimal objective is $\frac{\rho^{(\star)}\Phi(\overline{x})}{\overline{x}}\geq \Prob{E}$. Moreover, observe that $\overline{x}$ always exists and $\Phi(x)/\overline{x}\in (0,1)$. The search of $\overline{x}$ can be simplified by exploiting the same approach used in~\cite{Pellegrina_2021} that leads to 
\begin{align*}
	&\Prob{E}\leq\sup_{x\in (0,\min\{\hat{x}_1,\hat{x}_2\})}\left\{\frac{\rho^{(\star)}\eta(x)}{x}\right\}\leq \sup_{x\in (0,\hat{x})}\left\{\frac{\rho^{(\star)}\gamma(g(x),s,\varepsilon)}{x}\right\} &
\end{align*}
Setting $s\geq \sup_{(0,\hat{x})}\left\{\ln(\frac{2\rho^{(\star)}}{x\delta})/(g(x)h(\frac{\varepsilon}{g(x)}))\right\}$ it holds that $\Prob{E}\leq \delta$. In order to approximate $s$, the r.h.s. of the equation can be computed using a numerical procedure~\cite{Pellegrina_2021,Pellegrina_2023} obtaining the following approximation: $$s\approx\frac{2\hat{v}+\frac{2}{3}\varepsilon}{\varepsilon^2}\left(\ln\left(\frac{2\rho^{(\star)}}{\hat{v}}\right)+\ln\left(\frac{1}{\delta}\right)\right)\in \bigO\left(\frac{\hat{v}+\varepsilon}{\varepsilon^2}\ln\left(\frac{\rho^{(\star)}}{\delta\hat{v}}\right)\right)$$
\end{proof}
Moreover, given Lemma~\ref{lemma:avg_dist_edges}, we have the following corollary for the $(\star)$-temporal edge betweenness:
\begin{corollary}\label{thm::mcera_ss_edges}
	Let $\calF= \{f(e), e\in E\}$ be a set of function from a domain $\calD$ to $[0,1]$. Let $f(e)$ be a function such that $\Expec{f(e)} = \gbet{e}$. Define $\hat{v} \in (0,1/4]$ and $\Psi^{(\star)}\geq 0$ such that
	\begin{align*}
		& \max_{e\in E} \texttt{Var}(f(e))\leq \hat{v}&\text{and}&& \sum_{e\in E}\gbet{e}\leq \Psi^{(\star)}&
	\end{align*}
	fix $\varepsilon,\delta\in (0,1)$, and let $\calS$ be an $i.i.d.$ sample taken from $\calD$ of size
	\begin{align*}
		|\calS|\in\bigO\left(\frac{\hat{v}+\varepsilon}{\varepsilon^2}\ln\left(\frac{\Psi^{(\star)}}{\delta\hat{v}}\right)\right)
	\end{align*}
	It holds that $SD(\calF,\calS)\leq \varepsilon$ with probability $1-\delta$ over $\calS$.
\end{corollary}
We make use of the following theorem to prove Theorem~\ref{thm:sup_dev_rade}.
\begin{theorem}[\cite{Boucheron_2013}]~\label{thm:boucheron}
	With probability at least $1-\delta$ over $\calS$, it holds
	\begin{align*}
		R(\calF,r)\leq R(\calF,\calS)+ \sqrt{\left(\frac{\ln(1/\delta)}{r}\right)+\frac{2\ln(1/\delta)R(\calF,\calS)}{r}}+\frac{\ln(1/\delta)}{r}
	\end{align*}
\end{theorem}
\paragraph{\textbf{Proof of Theorem~\ref{thm:sup_dev_rade}}}
\begin{proof}
	Define the following events: $E_1 = `` R(\calF,\calS)> \tilde{R}\text{''},E_2 = `` R(\calF,s)> \tilde{R}\text{''}, E_3 = ``\sup_{f\in\calF}\{\left|\mu_\calS(f)-\mu_\pi(f)\right|\}>\varepsilon_{\calF}\text{''}$, and $E = ``SD(\calF,\calS)> \varepsilon_{\calF}\text{''}$. Moreover, by applying the union bound on these events we obtain $\Prob{E}\leq E_1+E_2+E_3 $. Now we can upper bound the probabilities of the single events by applying the \emph{symmetrization lemma} (Theorem 14.20 in~\cite{Mitzenmacher_2017}), Theorem 2.3 in~\cite{Bousquet_2002} and Theorem~\ref{thm:boucheron} replacing $\delta/4$ in the equations:
	$\Prob{E_1}\leq \delta/4$ follows from Equation (1) in Theorem~\ref{thm:sup_dev_rade} by replacing $\delta$ with $\delta/4$; $\Prob{E_2}\leq\delta/4$ follows from Theorem~\ref{thm:boucheron}; and, $\Prob{E_3}\leq \delta/4$ follows after using the symmetrization lemma and twice Theorem 2.3 in~\cite{Bousquet_2002}. Moreover, the event $E$ is true with probability at most $\delta$.
\end{proof}

\paragraph{\textbf{Proof of Theorem~\ref{thm:eps_delta_guar}}.}
\begin{proof}
	Each coordinate in $\agbet{v},v\in V$ is a sample mean (over a sample $\calS$ of size $r$) of a specific function associated to an unbiased estimator for $\gbet{v}$. Algorithm~\ref{algo:prog_scheme} stops if the number of the drawn samples is at least $\omega$ or if the supremum deviation bound $\xi$ is at most $\varepsilon$. In other words it stops when the $SD(\calF,\calS)\leq \varepsilon$ and in both cases this is guaranteed (by Theorem~\ref{thm::mcera_ss} or Theorem~\ref{thm:sup_dev_rade}) to happen with probability $1-\delta$.
\end{proof}
\paragraph{\textbf{Proof of Theorem~\ref{thm:mantra_running_time}}.}
\begin{proof}
	{\sc MANTRA} performs $r$ truncated $(\star)$-TBFS visits and regularly check the stopping condition until convergence. The running times of the temporal traversals depend on the type of path optimality we consider. Moreover, each TBFS requires $\bigO(n\cdot |T|\cdot\log\left(n\cdot |T|\right))$~\cite{Wu_2014,Buss_2020} to compute the shortest (foremost) temporal betweenness and $\bigO(|\calE|\cdot\log|\calE|)$ to compute the prefix-foremost temporal betweenness. Furthermore, to compute and check the stopping condition of the progressive sampler we need roughly linear time in $n$. Thus {\sc MANTRA}'s running time is $\bigO(r\cdot n\cdot |T|\cdot \log (n\cdot |T|)) = \tilde{\bigO}(r\cdot n\cdot |T|)$ for the shortest and shortest foremost temporal betweenness and $\bigO(r\cdot |\calE|\cdot \log |\calE|)=\tilde{\bigO}(r\cdot |\calE|)$ for the prefix-foremost temporal betweenness. Finally, we observe that the space required by MANTRA is $\bigO(n+|\calE|+c\cdot n)$ observe that $c$ is a fixed constant ($c=25$), thus the overall needed space is $\bigO(n+|\calE|)$.
\end{proof}

\section{Other estimators for the $(\star)$-temporal betweenness centrality}
In this section we present other two unbiased estimators for the $(\star)$-temporal betweenness centrality.
\paragraph{The RTB Estimator.}\label{apx:other_estimators}
We define the \emph{\underline{R}andom \underline{T}emporal \underline{B}etweenness} estimator (\ssbe). An intuitive technique to obtain an approximation of the $(\star)$-temporal betweenness centrality of a temporal graph $\calG$ is to run the exact temporal betweenness algorithm on a subset $S$ of nodes selected uniformly at random from $V$. Thus, in this case, the sampling space $\calD_{\ssbe}$ is the set $V$ of vertices in $\calG$, and the distribution $\pi_{\ssbe}$ is uniform over this set. The family $\calF_{\ssbe} = \{\afunc{v|s}{\ssbe}:v\in V\}$, contains one function $\afunc{v|s}{\ssbe}$ for each vertex $v$, defined as:
\begin{align}
	\afunc{v|s}{{\ssbe}} = \frac{1}{n-1}\cdot\sum_{\substack{z\in V\\ z\neq s}}\frac{\sigma_{sz}^{(\star)}(v)}{\sigma_{sz}^{(\star)}} \in [0,1]
\end{align}
It follows that
\begin{lemma}\label{lemma:ssbe}
	The \ssbe~is an unbiased estimator of the $(\star)$-temporal betweenness centrality.
\end{lemma}
\begin{proof}\label{proof:ssbe}
	\begin{align*}
		&\Expec{\afunc{v|s}{\ssbe}} = \sum_{s\in V} \Prob{s}\cdot \afunc{v|s}{\ssbe} = \sum_{s\in V}\frac{1}{n}\left(\frac{1}{n-1}\sum_{z\neq s\neq v}{\frac{\sigma^{(\star)}_{sz}(v)}{\sigma^{(\star)}_{sz}}} \right)&
	\end{align*}
\end{proof}
The function $\afunc{v|s}{\ssbe}$ is computed by performing a full $(\star)$-temporal breadth first search visit ($(\star)$-TBFS for short) from $s$, and then backtracking along the temporal directed acyclic graph as in the exacts algorithms \cite{Buss_2020}. Moreover, the following lemma holds:
The \ssbe~ framework computes \emph{all} the sets $\sopt{sz}$ from the sampled vertex $s$ to \emph{all} other vertices $z\in V$ using a full $(\star)$-TBFS. Moreover, in a worst-case scenario this algorithm could touch all the temporal edges in the temporal graph at \emph{every sample} making the estimation process slow. As for the static case \cite{Brandes_2007,Jacob_2004}, this algorithm does not scale well as the temporal network size increases.
\paragraph{The Temporal Riondato and Kornaropoulos estimator.}
We extend the estimator for static graphs by Riondato and Kornaropoulos in \cite{Riondato_2014} to the temporal setting. The algorithm, (1) computes the set $\sopt{sz}$ as \onbra; (2) randomly selects a \optpath $tp_{sz}$ from $\sopt{sz}$; and, (3) increases by $\frac{1}{r}$ the temporal betweenness of each vertex $v$ in \textbf{Int}(tp) (where $r$ is the sample size).
The procedure to select a random temporal path from $\sopt{sz}$ is inspired by the dependencies accumulation  to compute the \emph{exact} temporal betweenness scores by Bu{\ss} et al.\cite{Buss_2020}. Let $s$ and $z$ be the vertices sampled by our algorithm. We assume that $s$ and $z$ are temporally connected, otherwise the only option is to select the empty temporal path $tp_\emptyset$. Given the set $\sopt{sz}$ of all the \optpaths from $s$ to $z$, first we notice that the truncated $(\star)$-TBFS from $s$ to $z$ produces a time respecting tree from the vertex appearance $(s,0)$ to all the vertex appearances of the type $(z,t_z)$ for some $t_z$. 
Let $tp^*$ be the sampled \optpath we build \emph{backwards} starting from one of the temporal endpoints of the type $(z,t_z)$ for some $t_z$. First, we sample such $(z,t_z)$ as follows: a vertex appearance $(z,t_z)$ is sampled with probability $ \sigma_{sz}^{t_z}/(\sum_t \sigma_{sz}^{t}) = \sigma_{sz}^{t_z}/\sigma_{sz}$, where $\sigma_{sz}^{t}$ is the number of \optpaths reaching $z$ from $s$ at time $t$. Assume that $(z,t_z)$ was put in the sampled path $tp^*$, i.e., $tp^* = \{(z,t_z)\}$.
Now we proceed by sampling one of the temporal predecessors $(w,t_w)$ in the \emph{temporal predecessors} set $P(z,t_z)$ with probability $\sigma_{sw}^{t_w}/(\sum_{(x,t)\in P(z,t_z)} \sigma_{sx}^{t})$. After putting the sampled vertex appearance, let us assume $(w,t_w)$, in $tp^*$ we iterate the same process through the predecessors of $(w,t_w)$ until we reach $(s,0)$.
\begin{theorem}\label{thm:path_sampler}
	Let $tp^*_{sz}\in\sopt{sz}$ be the $(\star)$-temporal path sampled using the above procedure. Then, the probability of sampling $tp^*_{sz}$ is $\Prob{tp^*} =\frac{1}{\sigma^{(\star)}_{sz}}$
\end{theorem}
\begin{proof}
	Let $\sigma_{sz} = \sigma_{sz}^{(\star)}$. The probability of getting such $tp^*$ using the  aforementioned temporal path sampling technique is:
	\begin{align*}
		&\Prob{tp^*} = \frac{\sigma_{sz}^{t_z}}{\sum_t \sigma_{sz}^{t}} \cdot  \frac{\sigma_{sw_{k-1}}^{t_{w_{k-1}}}}{\sum_{(x,t)\in P(z,t_z)} \sigma_{sx}^{t}}\cdot \frac{\sigma_{sw_{k-2}}^{t_{w_{k-2}}}}{\sum_{(x,t)\in P(w_{k-1},t_{k-1})} \sigma_{sx}^{t}}  \cdots &\\& \frac{\sigma_{sw_{1}}^{t_{w_{1}}}}{\sum_{(x,t)\in P(w_{2},t_{2})} \sigma_{sx}^{t}}\cdot \frac{1}{\sum_{(x,t)\in P(w_1,t_1)} \sigma_{sx}^{t}}
	\end{align*}
	Observe that $\sigma_{sw}^{t_{w}} = \sum_{(x,t)\in P(w,t_w)} \sigma_{sx}^{t}$ and that $\sum_t \sigma_{sz}^{t} = \sigma_{sz}$. Thus, the formula can be rewritten as follows:
	\begin{align*}
		\Prob{tp^*} = \frac{\sigma_{sz}^{t_z}}{\sigma_{sz}} \cdot  \frac{\sigma_{sw_{k-1}}^{t_{w_{k-1}}}}{\sigma_{sz}^{t_z}} \cdots \frac{1}{\sigma_{sw_{1}}^{t_{w_{1}}}} = \frac{1}{\sigma_{sz}}
	\end{align*}
	and the fact that (if the temporal graph has no self loop) for $(w_1,t_{w_1})$, which is a temporal neighbor of $(s,0)$, $\sigma_{s w_1} = 1$.
\end{proof}
Observe that each $tp_{sz}\in\dopt_\calG$ is sampled according to the function $\pi_\srtp(tp_{sz}) = \frac{1}{n(n-1)}\frac{1}{\sigma^{(\star)}_{sz}}$ which (according to Theorem \ref{thm:prob_dist}) is a valid probability distribution over $\calD_{\srtp} = \dopt_\calG$.
\begin{theorem}\label{thm:prob_dist}
	The function $\pi_\srtp(tp_{sz})$, for each $tp_{sz}\in\calD_{\srtp}$, is a valid probability distribution.
\end{theorem}
\begin{proof}
	Let $\sopt{sz}$ be the set of $(\star)$-optimal temporal paths from $s$ to $z$ where $s\neq z$. Then,
	\begin{align*}
		&\sum_{tp_{sz}\in \calD_{\srtp}}\pi(tp_{sz})= \sum_{tp_{sz}\in \calD_{\srtp}}\frac{1}{n(n-1)}\frac{1}{\sigma_{sz}^{(\star)}} =&\\
		&\sum_{s\in V}\sum_{\substack{z\in V\\ s\neq z}}\sum_{tp_{sz}\in \sopt{sz}}\frac{1}{n(n-1)}\frac{1}{\sigma_{sz}^{(\star)}} = \sum_{s\in V}\sum_{\substack{z\in V\\ s\neq z}}\frac{1}{n(n-1)}\frac{\sigma_{sz}^{(\star)}}{\sigma_{sz}^{(\star)}}&\\&  = \frac{1}{n(n-1)}\sum_{s\in V}\sum_{\substack{z\in V\\ s\neq z}} 1 = \frac{1}{n(n-1)}\sum_{s\in V}(n-1) = 1
	\end{align*}
\end{proof}
For $tp_{sz}\in \calD_{\srtp}$, and for all $v\in V$ define the family of functions  $\calF_\srtp = \{\afuncc{v}{\srtp} : v\in V\}$ where $ \afunc{v|tp_{sz}}{\srtp}  = \mathds{1}\left[v\in \textbf{ Int}(tp_{sz})\right] $. Observe that
\begin{lemma}\label{lemma:rtp_unbiased}
	For $\afunc{v}{\srtp}\in\calF$ and for all $tp_{sz}\in  \calD_{\srtp}$, such that each $tp_{sz}$ is sampled according to the probability function $\pi(tp_{sz})$, then $
	\Expec{\afunc{v|tp_{sz}}{\srtp}} = \gbet{v}$.
\end{lemma}
\begin{proof}
	\begin{align*}
		&\CExpec{\afunc{v|tp_{sz}}{\srtp}}{tp_{sz}\in  \calD_{\srtp}} = \sum_{tp_{sz}\in\calD_{\srtp}} \pi(tp_{sz})\afunc{v|tp_{sz}}{\srtp} =&\\&\sum_{tp_{sz}\in \calD_{\srtp}} \frac{\afunc{v|tp_{sz}}{\srtp}}{n(n-1)\sigma_{sz}^{(\star)}}  = \frac{1}{n(n-1)} \sum_{\substack{s,z\in V\\ s\neq v\neq z}}\sum_{tp\in \sopt{sz}}  \frac{\mathds{1}\left[v\in \textbf{ Int}(tp)\right]}{\sigma^{(\star)}_{sz}}
		&\\ & =\frac{1}{n(n-1)} \sum_{\substack{s,z\in V\\ s\neq v\neq z}}\frac{\sigma^{(\star)}_{sz}(v)}{\sigma^{(\star)}_{sz}} &
	\end{align*}
\end{proof}

\section{Approximating the $(\star)$-temporal distance based metrics.}\label{apx::temporal_distance_metrics}
Given a temporal graph $\calG= (V,\calE)$, let $N(u,h) = \{v\in V : d(u,v)\leq h\}$ be the \emph{temporal ball} centered in $u$ at time $0$ of radius $h$. $N(u,h)$ is the set of all the nodes $v$ that can be reached by node $u$ starting at time $0$ by a $(\star)$-optimal temporal path. Now define $|N(h)| = |\{(u,v)\in V\times V : d(u,v)\leq h\}|$ as the overall pairs of nodes that can be reached by a $(\star)$-optimal path in $h$ steps. The $(\star)$-temporal diameter $D$ of a graph is the number of temporal edges in the longest $(\star)$-temporal path in the temporal graph. In terms of $N(h)$ we have:
\begin{align*}
	D^{(\star)} = \min_{h}\left(h:\sum_{u}|N(u,h)| = \sum_{u}|N(u,h+1)|\right)
\end{align*}
Alternatively, as for the static case, we can define the \emph{effective $(\star)$-temporal diameter} as the $\tau^{(th)}$ percentile $(\star)$-temporal path length between the nodes. We will use this quantity to provide an error bound for the approximation of $D^{(\star)}$. Let $\tau \in [0,1]$, then
\begin{align*}
	D_{\tau}^{(\star)} = \min_{h}\left(h:\frac{\sum_{u}|N(u,h)|}{\sum_{u}|N(u,D^{(\star)})|} \geq \tau\right)
\end{align*}
Finally, observe that given $N(u,h)$ for each $u\in V$ and $h\in\{0,\dots ,D\}$ we can define the average $(\star)$-temporal distance 
\begin{align*}
	&\Psi^{(\star)} = \frac{\sum_{\substack{u,v\in V\\ u\neq v}}\mathds{1}[u\rightsquigarrow v]\cdot d(u,v)}{\sum_{\substack{u,v\in V\\ u\neq v}}\mathds{1}[u\rightsquigarrow v]} = \frac{\sum_{u\in V}\sum_{h\in [0,D^{(\star)}]} (|N(u,h)-|N(u,h-1))\cdot h}{|N(D^{(\star)})|}&\\=& \frac{\sum_{s,z\in V}\sum_{e\in E}\mathds{1}{[e\in tp_{sz}]}}{|N(D^{(\star)})|}&
\end{align*}

Observe that $\Psi^{(\star)}$ and $\rho^{(\star)}$ are tightly related, 
\begin{align*}
	\rho^{(\star)} = \frac{\sum_{u\in V}\sum_{h\in (0,D^{(\star)})} (|N(u,h)-|N(u,h-1))\cdot h}{|N(D^{(\star)})|} \leq \Psi^{(\star)}
\end{align*}
Algorithm~\ref{algo:tdiam}, given a temporal graph $\calG= (V,\calE)$, a number of seeds $r$ and the temporal path optimality of interest $(\star)$ the algorithm performs a $(\star)$-TBFS and keeps track of the number of nodes encountered at each hop and the maximum number of hops performed. Once all the seed nodes have been processed it computes the overall set of reachable pairs and the average distance $\Psi^{(\star)}$, effective diameter $D_{\texttt{LB}}$, and the temporal connectivity rate $\zeta_\calG$. 

\input{pseudocodes/diameter_apx}

Algorithm \ref{algo:tdiam}, computes the $(\star)$-temporal distance-based metrics described above. Moreover, we can give a bound on the quality of the approximation produced by this approach.
\begin{theorem}
	Given a temporal graph $\calG = (V,\calE)$ and a sample of size $r = \Theta \left(\frac{\ln n}{\varepsilon^2}\right)$. Algorithm \ref{algo:tdiam} computes with probability $1-\frac{2}{n^2}$  the $(\star)$ temporal (effective) diameter with absolute error bounded by $\varepsilon\over \zeta(\calG)$, the temporal connectivity rate $\zeta(\calG)$ with absolute error bounded by $\varepsilon$, and the average $(\star)$-temporal length with absolute error bounded by $\frac{\varepsilon\cdot D}{\zeta(\calG)}$.
\end{theorem}
\begin{proof}
	Let $h$ be the $(\star)$-temporal effective diameter threshold, and 
	\begin{align*}
		X_i^h = \frac{n\cdot \sum_{\{u:d(v_i,u)\leq h\} } \mathds{1}[v_i\rightsquigarrow u]}{\sum_{\substack{u,v\in V\\ u\neq v}}\mathds{1}[u\rightsquigarrow v]} = \frac{n\cdot |N(v_i,h)|}{|N(D)|}
	\end{align*}	
	observe that $X_i^h \in [0,\frac{1}{\zeta}]$, and that $X_i^h$'s expected value is 
	\begin{align*}
		\Expec{X_i^h} = \sum_{v_i\in V}X_i^h\cdot \Prob{v_i} = \frac{ \sum_{v_i\in V}|N(v_i,h)|}{|N(D)|} = \frac{|N(h)|}{|N(D)|}
	\end{align*}
	Applying Hoeffding's inequality, with $\xi = \frac{\varepsilon}{\zeta}$, we can approximate $ \frac{|N(h)|}{|N(D)|}$ by $\frac{\sum_{i\in [r]}X_i^h}{r} = \frac{n\cdot \sum_{i\in [r]}|N(u,v_i)| }{r\cdot |N(D)|}$, and taking a sample of $r = {\ln n\over \varepsilon^2}$ leads an error of $\varepsilon\over\zeta$ with probability of at least $1-\frac{2}{n^2}$. Now we observe that the $(\star)$-temporal diameter can be defined in terms of the effective one by choosing $\tau = 1$, thus the bound holds also for the $(\star)$-temporal diameter. Next, define the random variable $X_i^D = \frac{|N(v_i,D)|}{n-1}$, and observe that $X_i^D \in [0,1]$. Observe that its expected value is exactly $\zeta(\calG)$, i.e., $\Expec{X_i^D} = \sum_{v_i\in V}\frac{1}{n}\frac{|N(v_i,D)|}{n-1} = \frac{|N(D)|}{n(n-1)} = \zeta(\calG)$. Again, applying Hoeffding's inequality, with $\xi = \varepsilon$, and taking $r = {\ln n\over\varepsilon^2}$ sample nodes, we have an error bound of $\varepsilon$ with high probability. Finally, define 
	\begin{align*}
		X_i = \frac{n\cdot \sum_{u\in V}\mathds{1}[v_i\rightsquigarrow u]\cdot d(v_i,u) }{\sum_{\substack{u,v\in V}}\mathds{1}[u\rightsquigarrow v]}
	\end{align*}
	Observe that $X_i\in [0,\frac{D}{\zeta(\calG)}]$, and that its expectation is the average shortest temporal distance
	\begin{align*}
		\Expec{X_i} = \sum_{v_i\in V} X_i \cdot \Prob{X_i} =  \frac{\sum_{\substack{v_i,u\in V}}\mathds{1}[v_i\rightsquigarrow u]\cdot d(v_i,u)}{\sum_{\substack{u,v\in V}}\mathds{1}[u\rightsquigarrow v]}
	\end{align*}
	Finally, by using Hoeffding's inequality with $\xi= \frac{\varepsilon D}{\zeta(\calG)}$ and setting $r = \frac{\log n}{\varepsilon^2}$ we obtain that the error is of at most $\frac{\varepsilon D}{\zeta(\calG)}$ with probability $1-\frac{2}{n^2}$.
\end{proof}
Observe that Algorithm~\ref{algo:tdiam} is general, it can be used to compute several notions of temporal diameter~\cite{Wu_2014,Michail_2016,Crescenzi_2019,Calamai_2022,Oettershagen_2022}. 
\section{Missing Experiments}~\label{apx:missing_exp}
\begin{figure}[htb!]
	\centering
	\caption{Experimental evaluation (over 10 runs) of the sampling algorithm using $256$ seed nodes.
		Comparison between temporal effective diameter and the connectivity rate for the Shortest (foremost) and Prefix-Foremost temporal path optimalities. Last two log-plots, are the running times comparisons between the exact and our approximation algorithm. }\label{fig:apx_charac_qts}
	\begin{subfigure}{0.48\textwidth}
		\centering
		\includegraphics[scale=0.4]{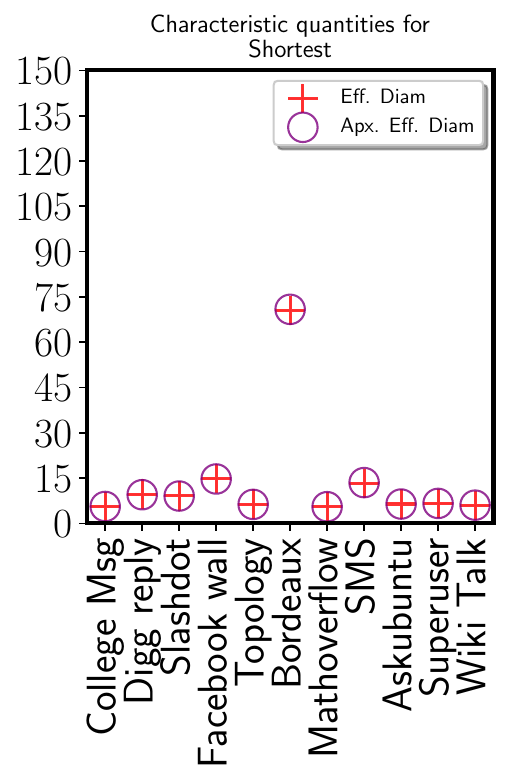}
	\end{subfigure}
	\begin{subfigure}{0.48\textwidth}
		\includegraphics[scale=0.4]{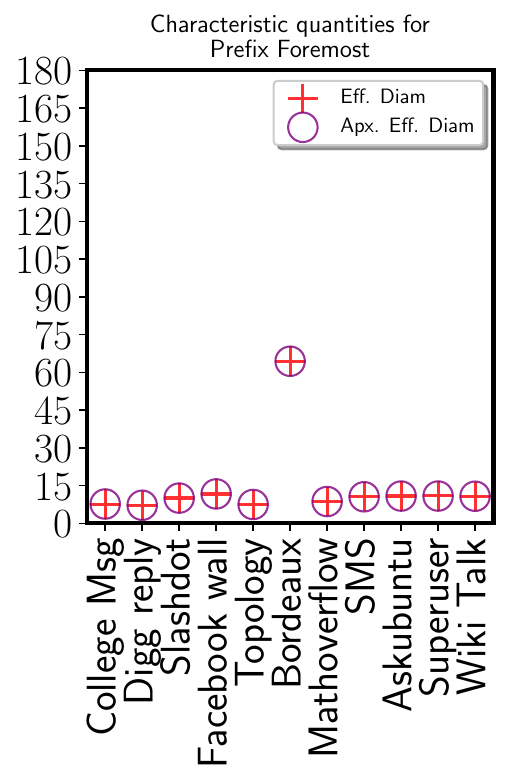}
	\end{subfigure}

	\begin{subfigure}{0.48\textwidth}
		\centering
		\includegraphics[scale=0.4]{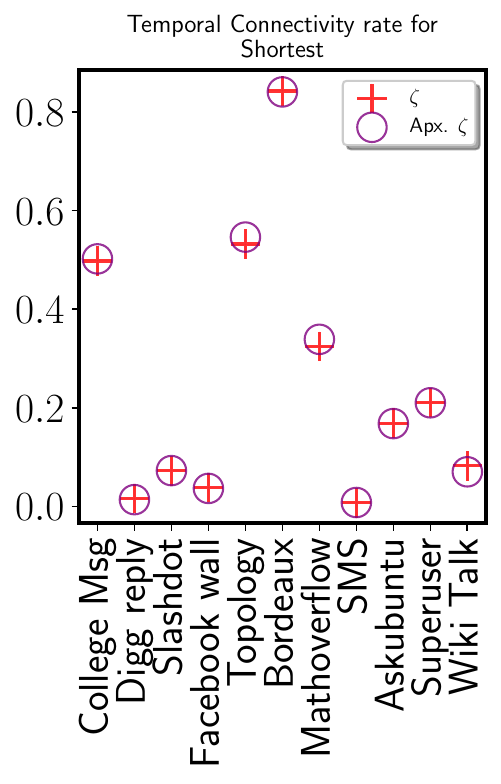}
	\end{subfigure}
	\begin{subfigure}{0.48\textwidth}
		\includegraphics[scale=0.4]{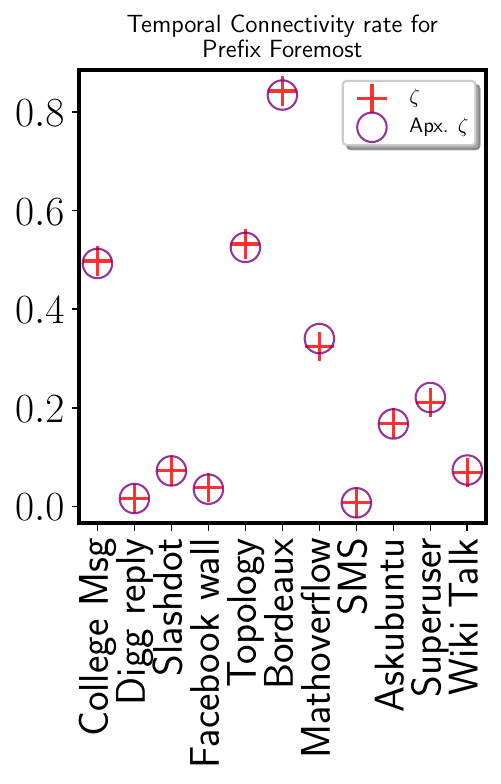}
	\end{subfigure}
	
	\begin{subfigure}{0.4\textwidth}
		\centering
		\includegraphics[scale=0.4]{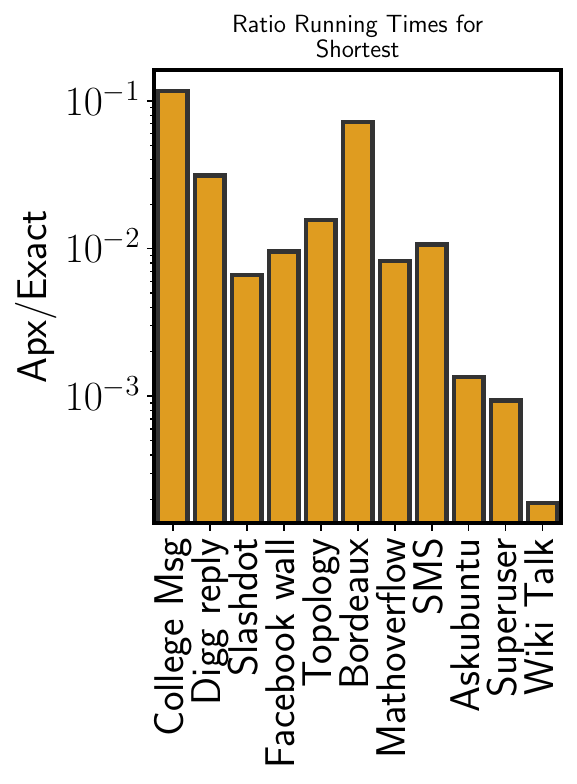}
	\end{subfigure}
	\begin{subfigure}{0.5\textwidth}
		\includegraphics[scale=0.4]{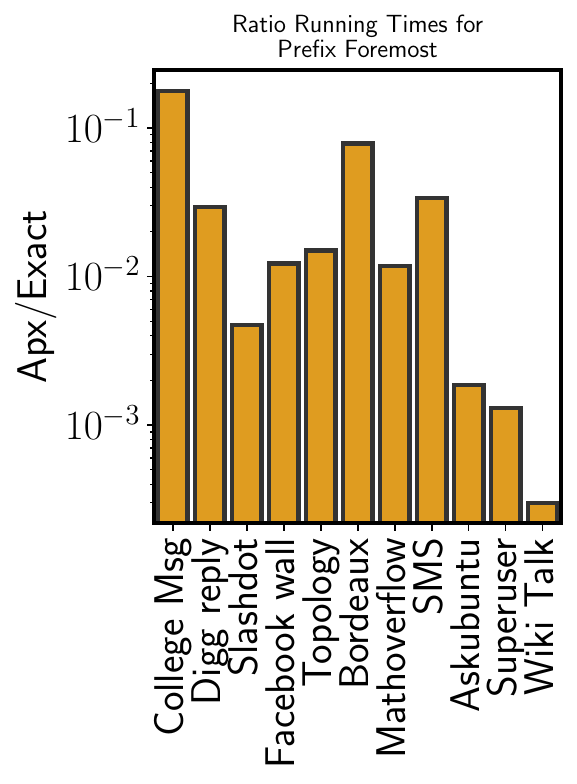}
	\end{subfigure}
\end{figure}

\begin{figure}[htb!]
	\caption{Experimental analysis for $\varepsilon\in\{0.01,0.007,0.005,0.001\}$ For the Shortest and Shortest-foremost temporal betweenness. Comparison between the running times  \textbf{(a-b)}, sample sizes \textbf{(c-d)}, and allocated memory \textbf{(e-f)} of ONBRA and MANTRA. \textbf{(g)} Supremum deviation of the absolute $\varepsilon$-approximation computed by MANTRA. }\label{fig:times_additional}
	\begin{subfigure}{0.2\textwidth}
		\centering
		\includegraphics[scale=0.135]{img/legend_2}
	\end{subfigure}
	
	\begin{subfigure}{0.5\textwidth}
		\includegraphics[scale=0.25]{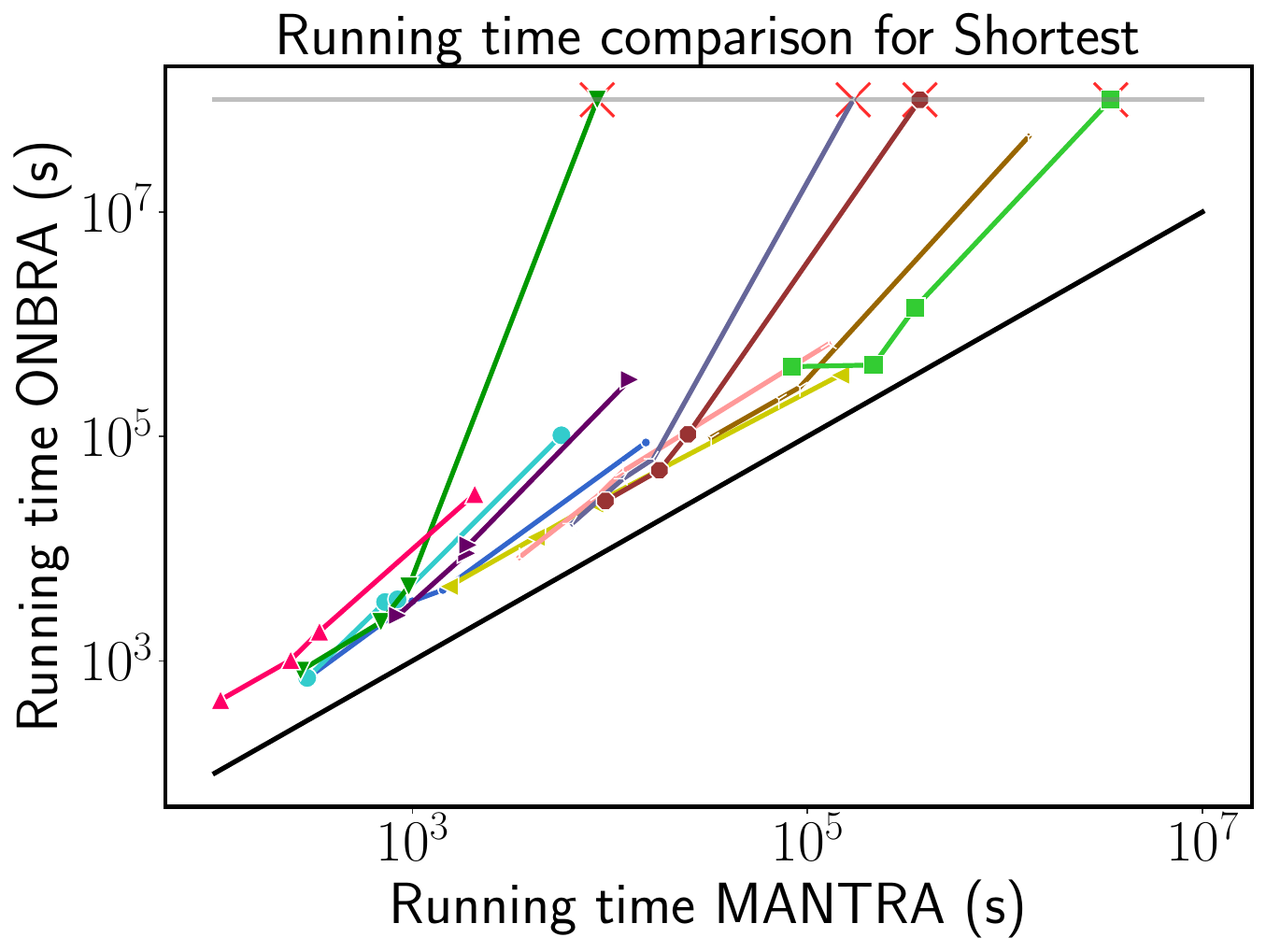}
		\caption{}\label{fig:times_a_sh}
	\end{subfigure}
	\begin{subfigure}{0.5\textwidth}
		\includegraphics[scale=0.25]{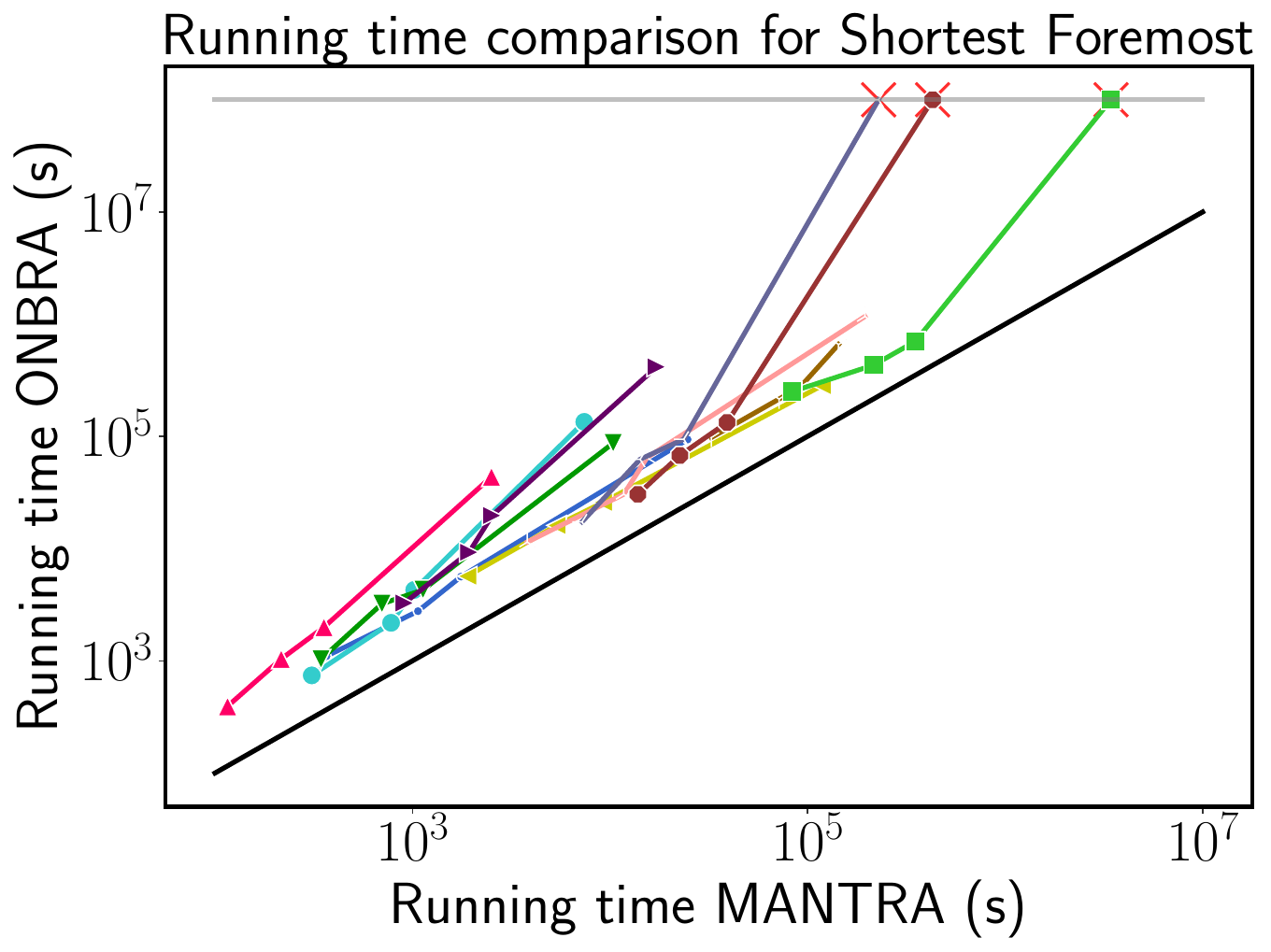}
		\caption{}\label{fig:times_a_sfm}
	\end{subfigure}
	
	\begin{subfigure}{0.5\textwidth}
		\includegraphics[scale=0.25]{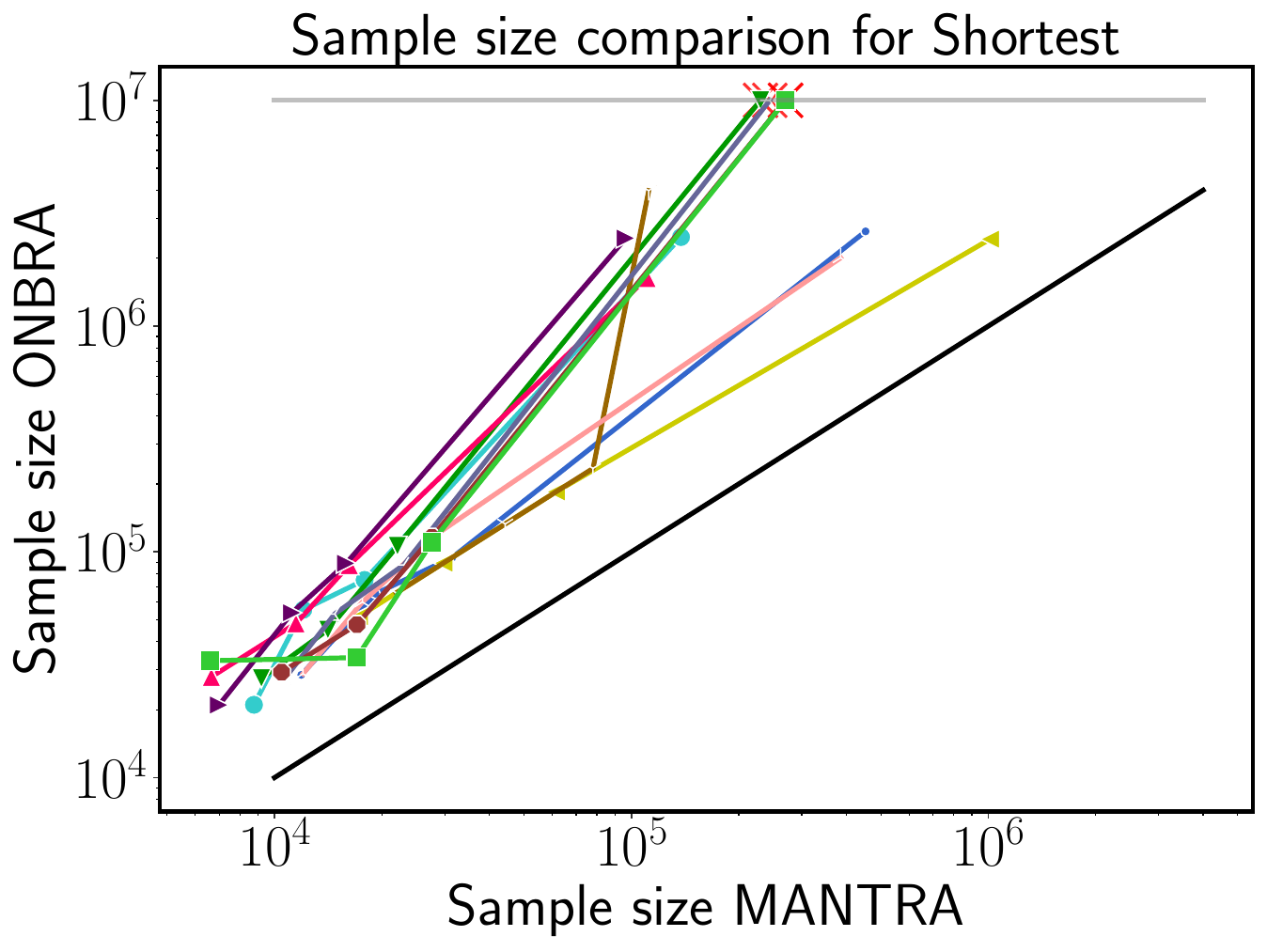}
		\caption{}\label{fig:ss_a_sh}
	\end{subfigure}
	\begin{subfigure}{0.5\textwidth}
		\includegraphics[scale=0.25]{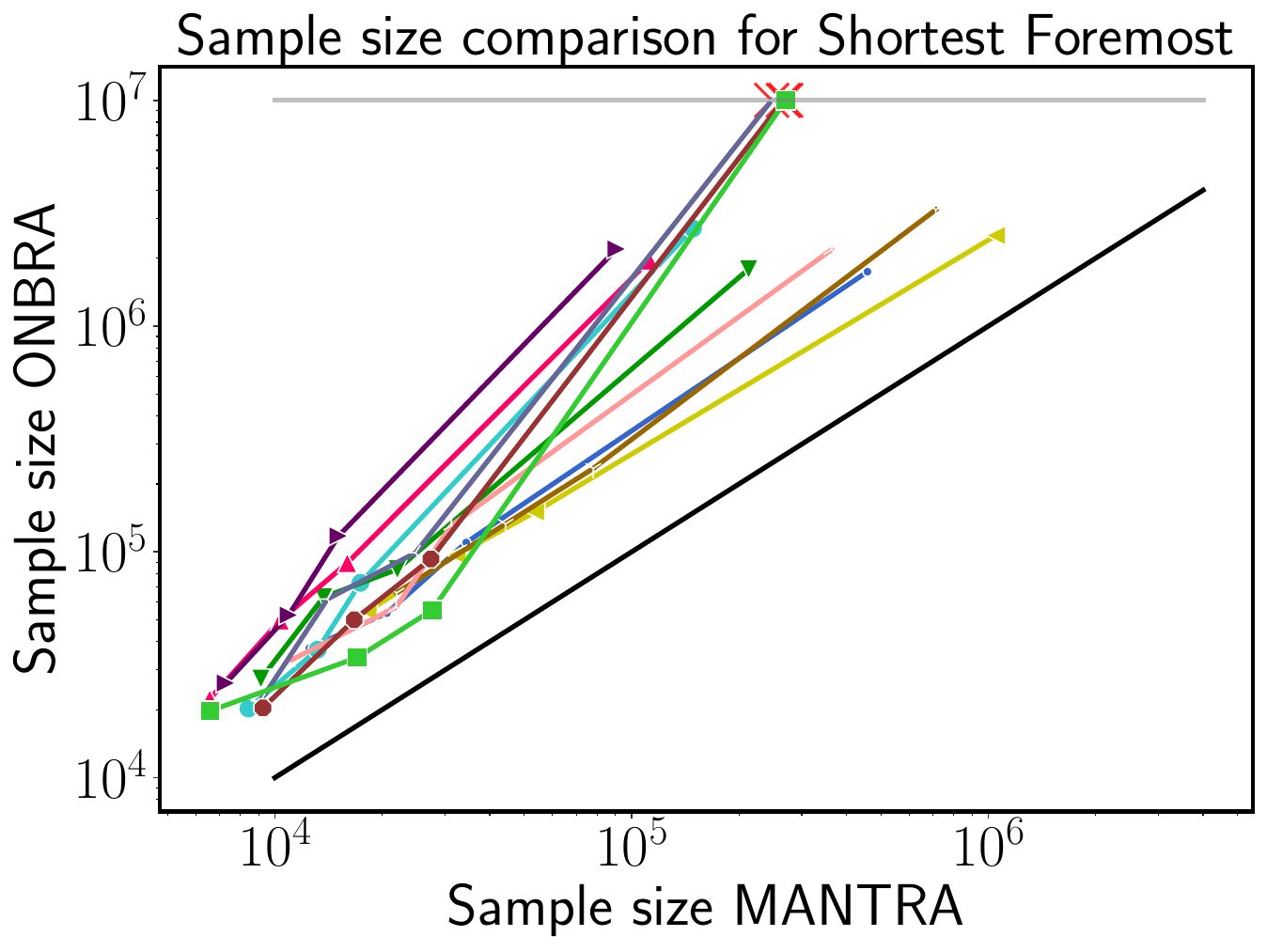}
		\caption{}\label{fig:ss_a_sfm}
	\end{subfigure}
	
	\begin{subfigure}{0.5\textwidth}
		\includegraphics[scale=0.25]{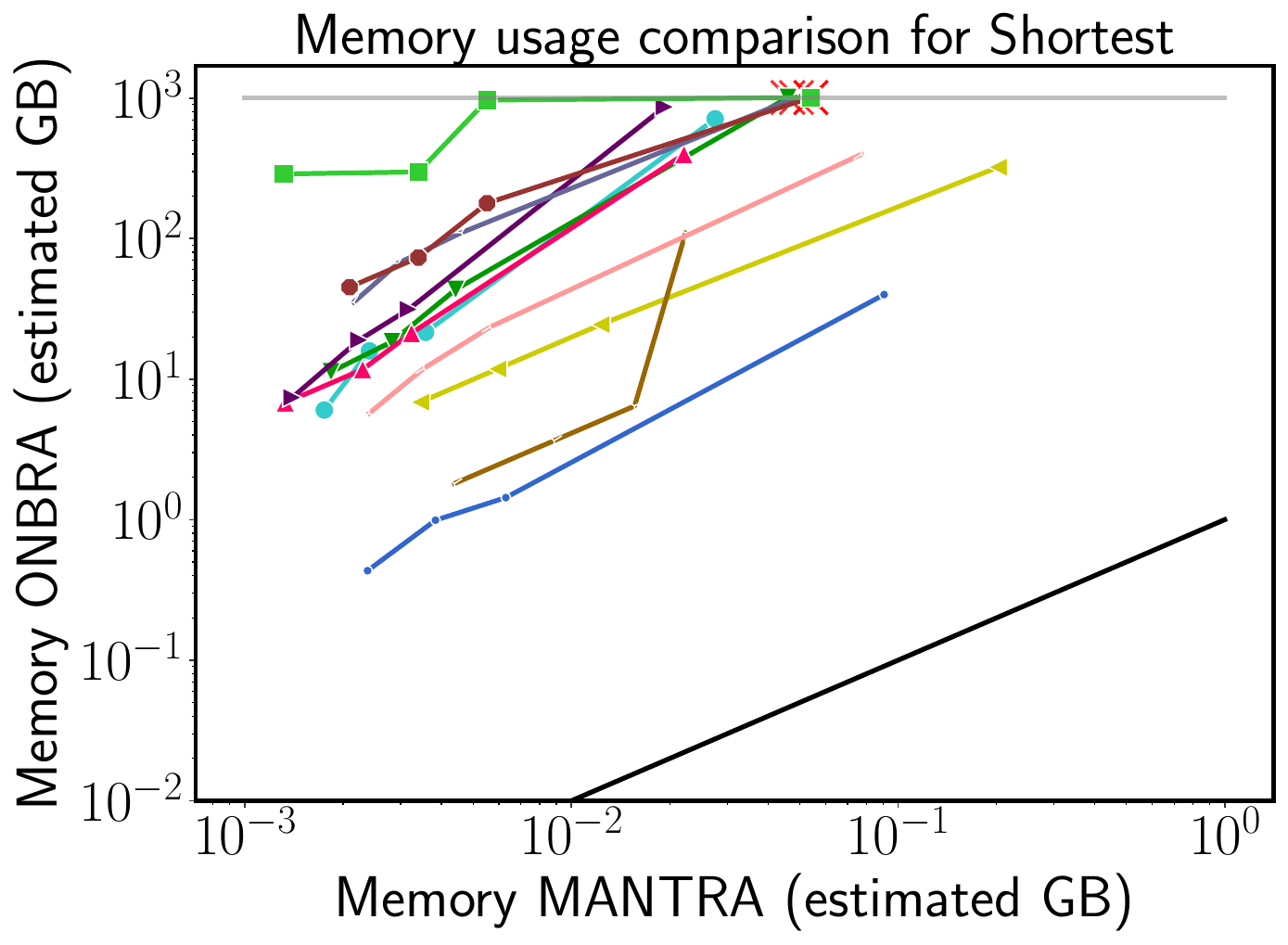}
		\caption{}\label{fig:space_a_sh}
	\end{subfigure}
		\begin{subfigure}{0.5\textwidth}
		\includegraphics[scale=0.25]{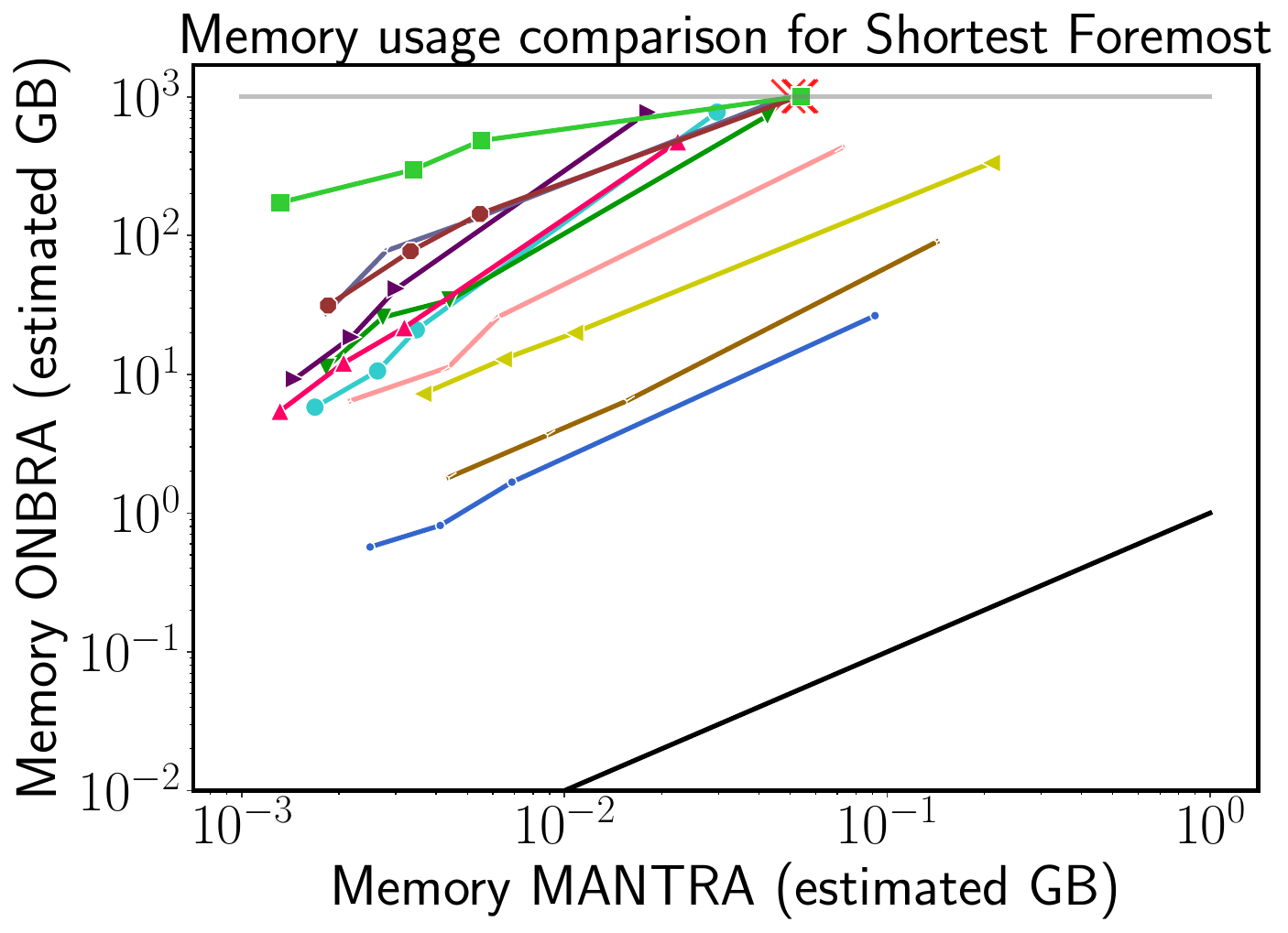}
		\caption{}\label{fig:space_a_sfm}
	\end{subfigure}
	
	\begin{subfigure}{0.5\textwidth}
		\includegraphics[scale=0.25]{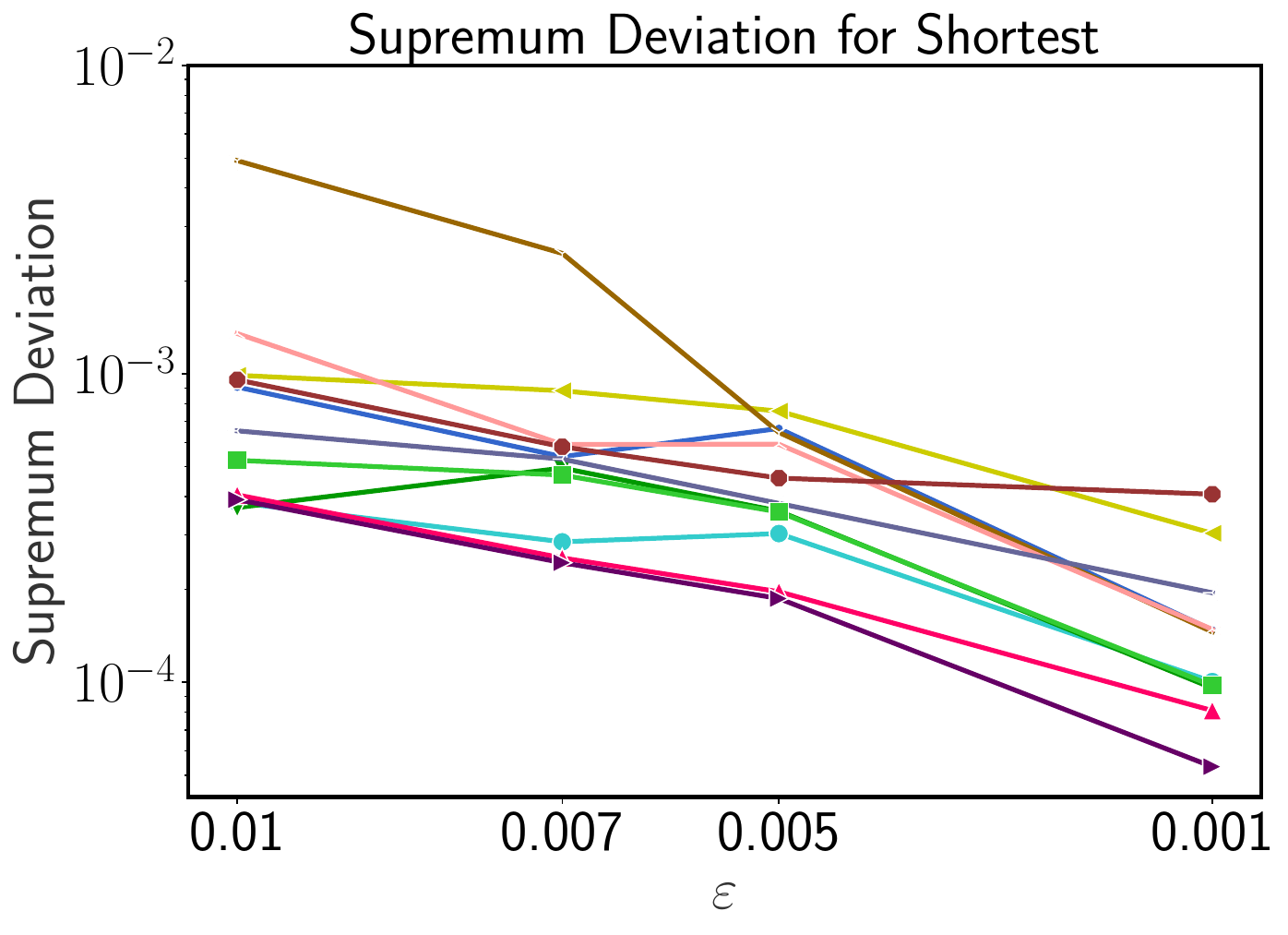}
		\caption{}\label{fig:sd_a_sh}
	\end{subfigure}	
	\begin{subfigure}{0.5\textwidth}
		\includegraphics[scale=0.25]{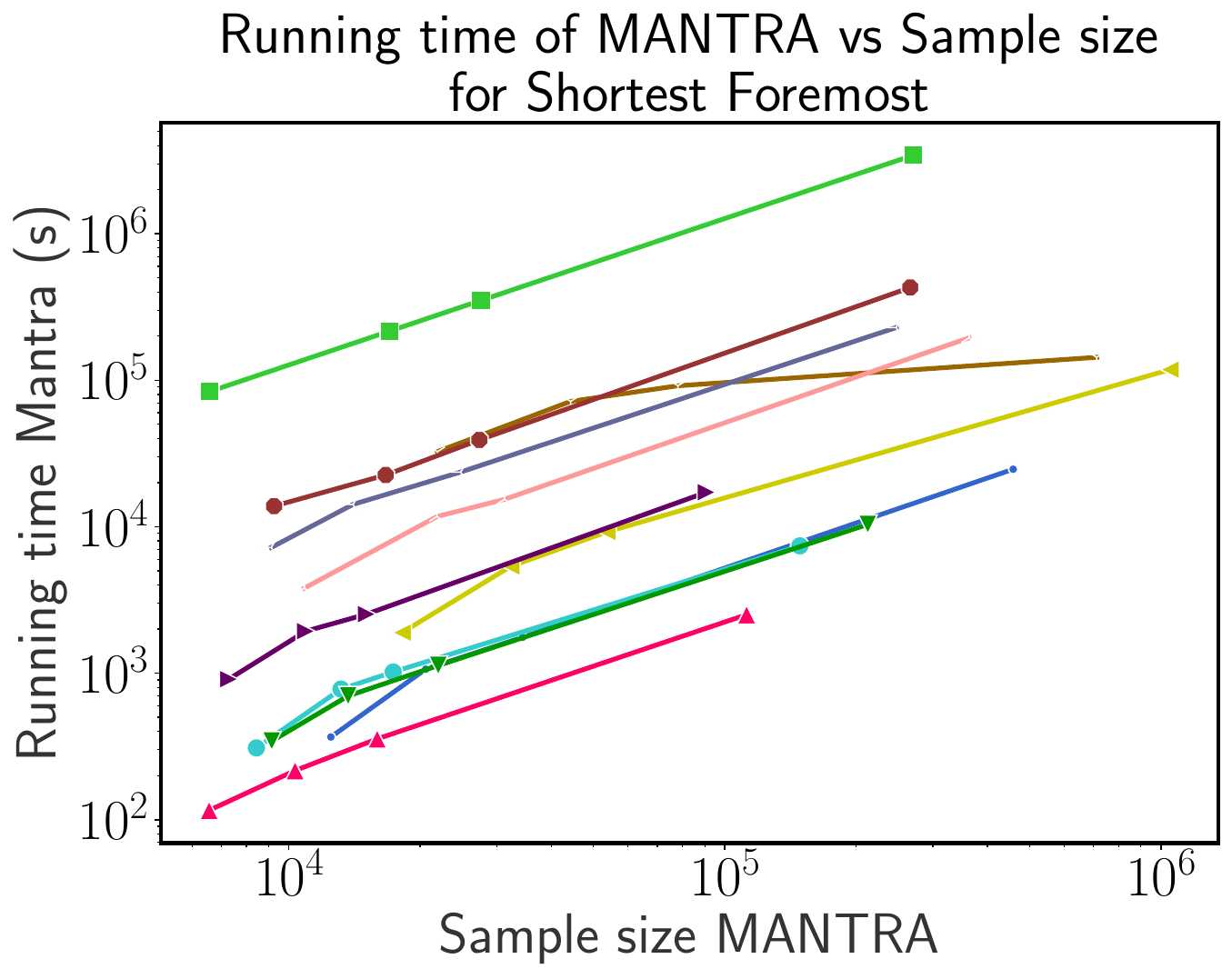}
		\caption{}\label{fig:runtime_vs_ss_sfm}
	\end{subfigure}

\end{figure}

\begin{figure}[htb!]
	\caption{Experimental analysis for $\varepsilon\in\{0.1,0.07,0.05\}$ For the Shortest and Shortest-foremost temporal betweenness. Comparison between the running times  \textbf{(a-b)}, sample sizes \textbf{(c-d)}, and allocated memory \textbf{(e-f)} of ONBRA and MANTRA. }\label{fig:times_small_additional}
	\begin{subfigure}{0.2\textwidth}
		\centering
		\includegraphics[scale=0.135]{img/legend_2}
	\end{subfigure}
	
	\begin{subfigure}{0.5\textwidth}
		\includegraphics[scale=0.25]{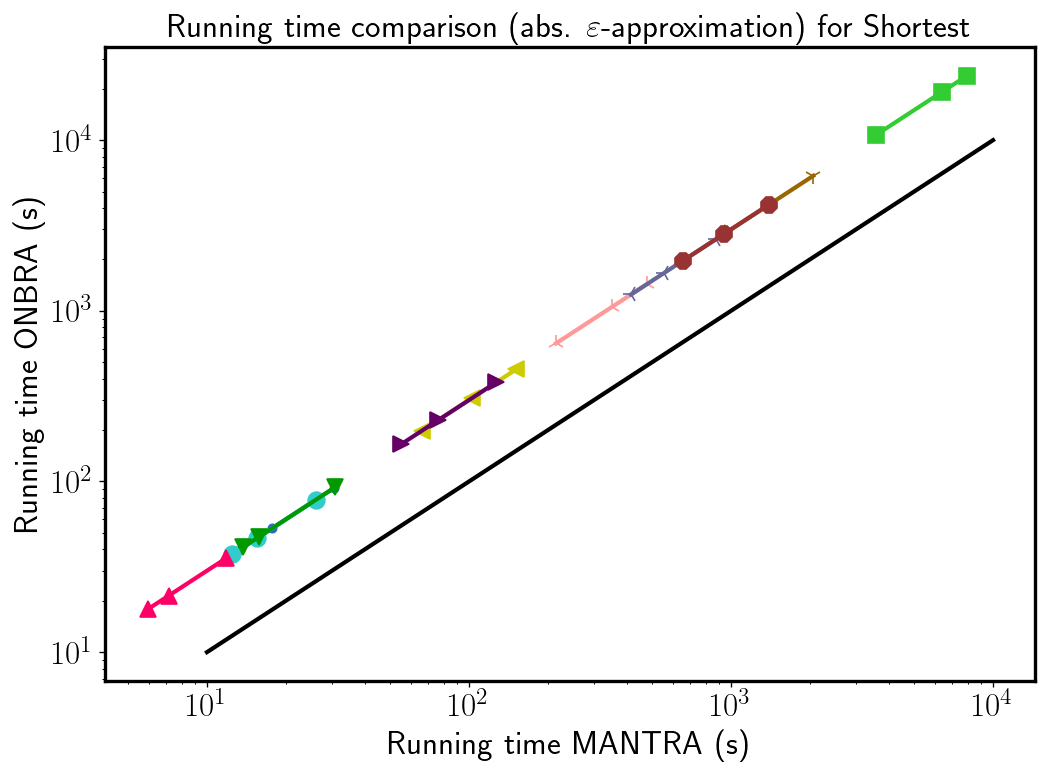}
		\caption{}\label{fig:times_a_small_sh}
	\end{subfigure}
	\begin{subfigure}{0.5\textwidth}
		\includegraphics[scale=0.25]{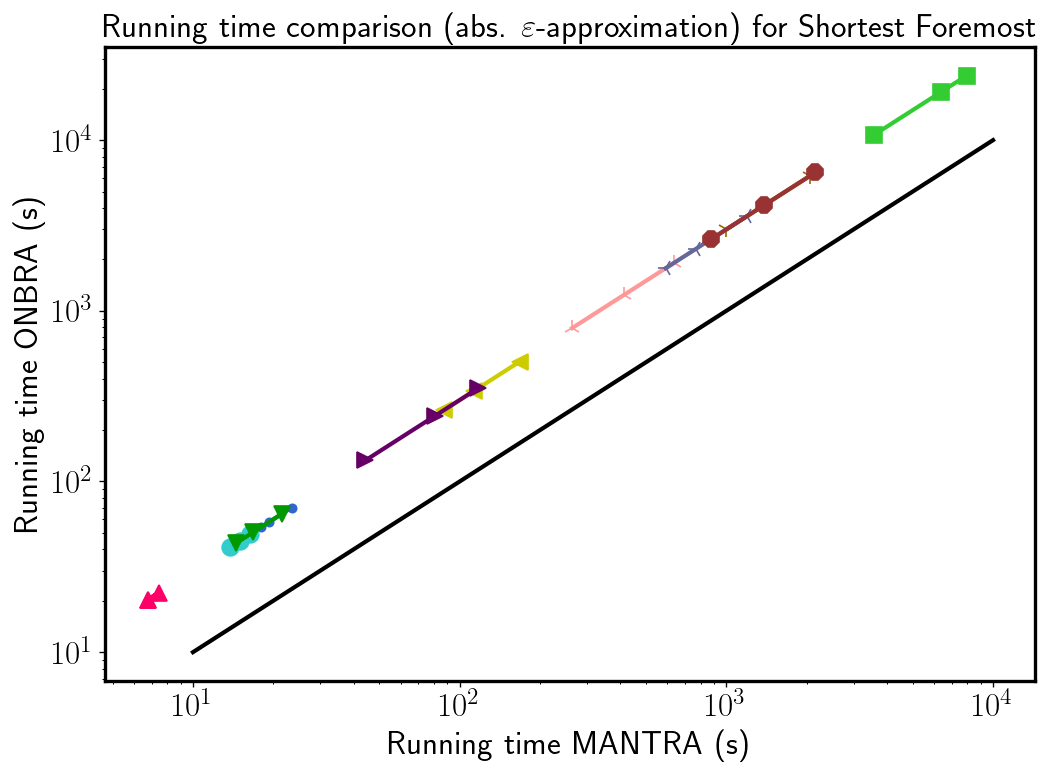}
		\caption{}\label{fig:times_a_small_sfm}
	\end{subfigure}
	
	\begin{subfigure}{0.5\textwidth}
		\includegraphics[scale=0.25]{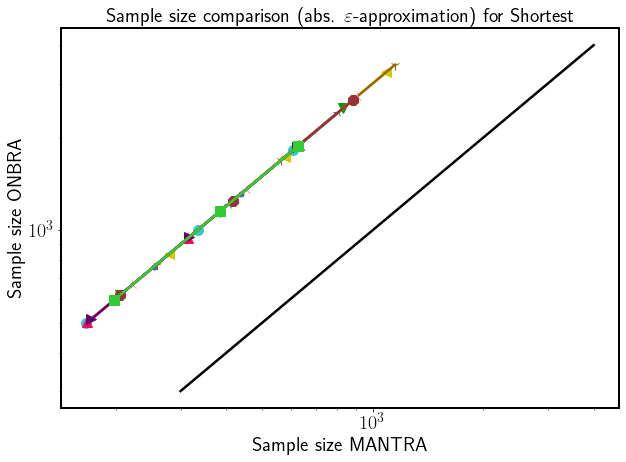}
		\caption{}\label{fig:ss_a_small_sh}
	\end{subfigure}
	\begin{subfigure}{0.5\textwidth}
		\includegraphics[scale=0.25]{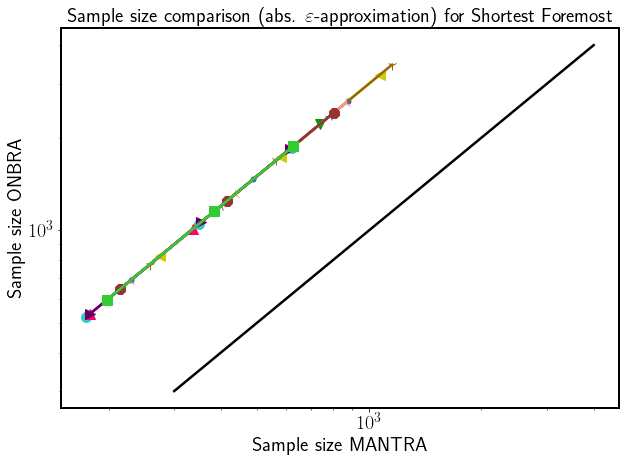}
		\caption{}\label{fig:ss_a_small_sfm}
	\end{subfigure}
	
	\begin{subfigure}{0.5\textwidth}
		\includegraphics[scale=0.25]{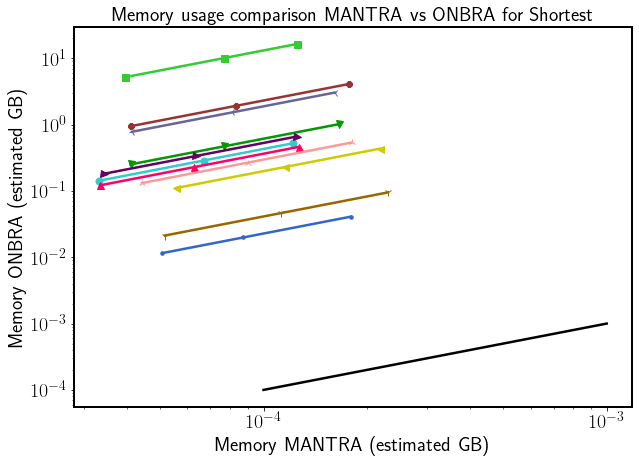}
		\caption{}\label{fig:space_small_a_sh}
	\end{subfigure}
	\begin{subfigure}{0.5\textwidth}
		\includegraphics[scale=0.25]{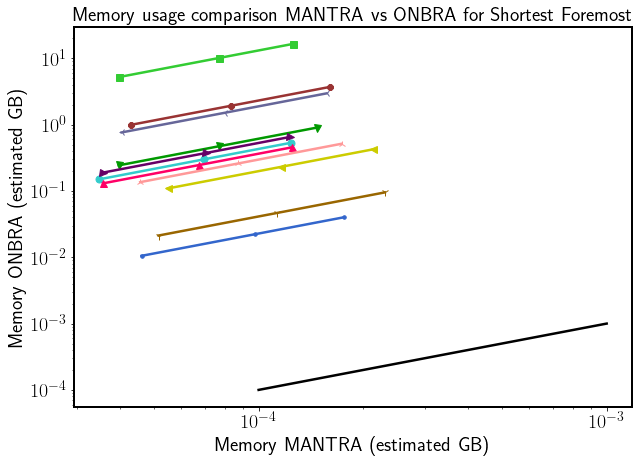}
		\caption{}\label{fig:space_a_small_sfm}
	\end{subfigure}
	
%
\end{figure}

\begin{figure}[htb!]
	\caption{Experimental analysis for $\varepsilon\in\{0.1,0.07,0.05\}$ For the Prefix-foremost temporal betweenness. Comparison between the running times  \textbf{(a)}, sample sizes \textbf{(b)}, and allocated memory \textbf{(c)} of ONBRA and MANTRA. }\label{fig:times_small_pfm_additional}
	\begin{subfigure}{0.2\textwidth}
		\centering
		\includegraphics[scale=0.135]{img/legend_2}
	\end{subfigure}
	
	\begin{subfigure}{0.5\textwidth}
		\includegraphics[scale=0.25]{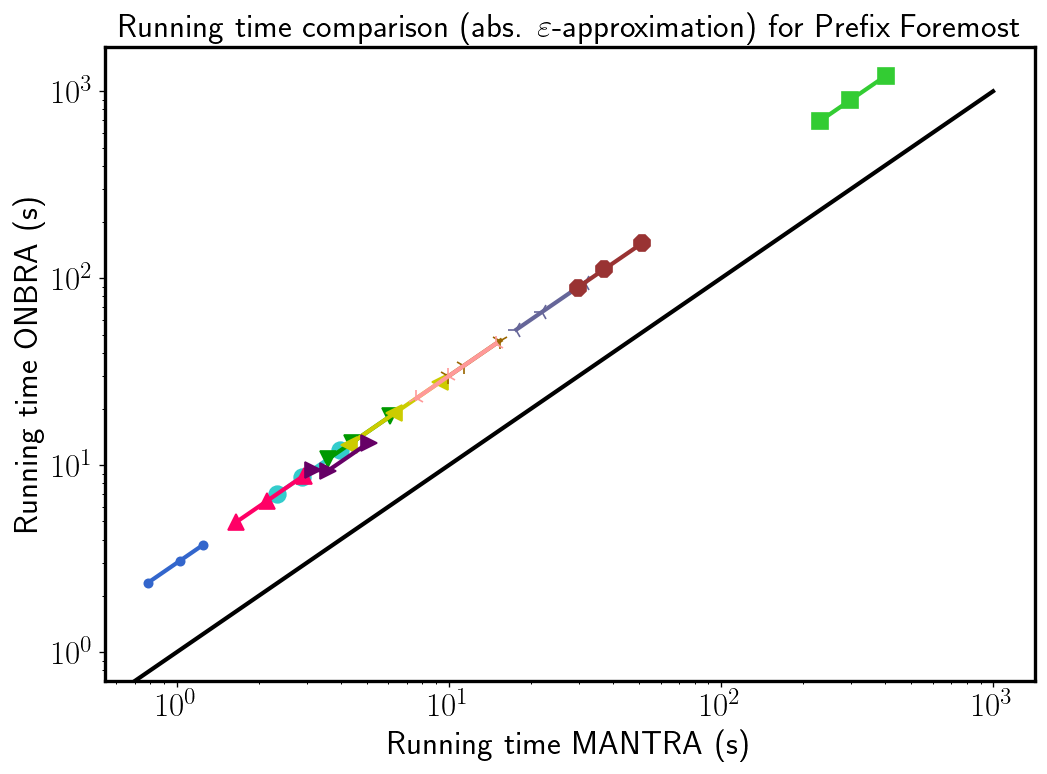}
		\caption{}\label{fig:times_a_small_pfm}
	\end{subfigure}
	\begin{subfigure}{0.5\textwidth}
		\includegraphics[scale=0.25]{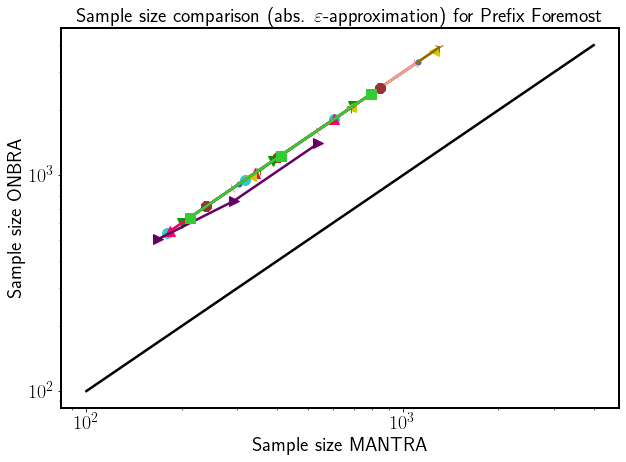}
		\caption{}\label{fig:ss_a_small_pfm}
	\end{subfigure}

	\begin{subfigure}{0.5\textwidth}
		\includegraphics[scale=0.25]{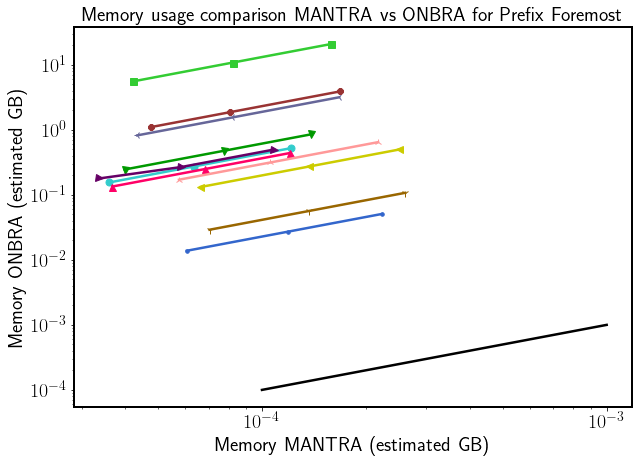}
		\caption{}\label{fig:space_small_a_pfm}
	\end{subfigure}

	%
\end{figure}

Figure~\ref{fig:apx_charac_qts} (in Section~\ref{apx:missing_exp}) and Figure~\ref{fig:diameter_and_spl} (in Section~\ref{sec::experiments}), show the average accuracy and execution times of Algorithm~\ref{algo:tdiam} (with $256$ seeds) compared to the time needed to retrieve exact values for the temporal networks in Table~\ref{tab::graphs}. We observe that our novel sampling algorithm provides sharp estimates of the exact values (temporal (effective) diameter, average path length, and connectivity rate). Moreover, the approximation process using $256$ seeds requires a negligible running time compared to the time needed to compute the exact characteristic quantities. This experimental analysis shows that the bootstrap phase of the MANTRA framework (that uses Algorithm~\ref{algo:tdiam} to approximate the characteristic quantities) provides very good estimates on the characteristic quantities that are used to compute the upper bound on the sample size. Moreover, we point out that Algorithm~\ref{algo:tdiam} could be used to provide fast approximation of the temporal Closeness Centrality (e.g.~\cite{Crescenzi_2020,Oettershagen_2022,Oettershagen_2023}). Finally, Figure~\ref{fig:times_additional}, Figure~\ref{fig:times_small_additional}, and Figure~\ref{fig:times_small_pfm_additional} show the experimental analysis for the shortest $(\star)$-temporal betweenness approximation for all the temporal path optimalieties and for the remaining values of $\varepsilon$.

%% file: pseudocodes/diameter_apx.tex
\begin{algorithm}
	
	\caption{$(\star)$-temporal (effective) diameter approximation}
	\label{algo:tdiam} 
	\KwData{Temporal graph $\calG,\varepsilon$, sample size $r$, and effective shortest temporal diameters's threshold $\tau$}
	\KwResult{Diameter lower bound $D_{\texttt{LB}}$, Effective diameter $D_\tau$, temporal connectivity rate $\zeta_\calG$ , average distance $\Psi$ }
	
	
	$R = [0,0,\dots ,0]$\tcp{Nr. of reach. pairs at each hop}
	
	$D_\texttt{LB} = 0;\Psi=0;D_\tau = -1$
	
	\For{$k = 0\texttt{ to }r-1$}{
		$u = \texttt{sample u.a.r. a node from } V$
		
		$R,D_{\texttt{LB}}=(\star)\texttt{-TemporalBFS}(\calG,u,R,D_{\texttt{LB}}) $ \tcp*{Update $R$ and $D_{\texttt{LB}}$}

%
%
%
%
%
%
%
%
%
%
%
%
%
	}

	\For{$h =0 \texttt{ to }D_\texttt{LB}-1 $}{

		$R[h] = \frac{n}{r} \cdot R[h] $
		
		\tcp{$R[-1]$ threaded as 0 when $h=0$}
		$\Psi =\Psi+ (R[h]-R[h-1])\cdot h$

	}
%
%

	$\Psi = \Psi/R[D_\texttt{LB}]$; $D_\tau = \min_h\left(h: \frac{R[h] }{R[D_\texttt{LB}]}\geq \tau \right)$; $\zeta_\calG = \frac{R[D_\texttt{LB}]}{n(n-1)}$
	
	\Return{$D_\texttt{LB},D_\tau,\zeta_\calG ,\Psi$}

\end{algorithm}